\documentclass[%
 reprint,
 amsmath,amssymb,
 aps,
 prd,
floatfix,
]{revtex4-2}

\usepackage{xspace}
\usepackage{subcaption}
\usepackage{multirow}
\usepackage{graphicx}
\usepackage{dcolumn}
\usepackage{bm}
\usepackage{hyperref}
\hypersetup{hidelinks}
\usepackage{dsfont}
\usepackage{comment}
\usepackage{xcolor}

\usepackage[normalem]{ulem}
\usepackage{soul}
\usepackage{float}
\usepackage{placeins}

\AtBeginDocument{\raggedbottom}
\begin{document}

\title{LISA Double White Dwarfs in Triple Systems}

\author{Zhijun Wang$^{1}$}
\author{Guoliang L\"{u}$^{1}$}%
\author{Chunhua Zhu$^{1}$}
\email{chunhuazhu@sina.cn}
\author{Xizhen Lu$^{1}$}
\author{Sufen Guo$^{1}$}
\author{Helei Liu$^{1}$}
\author{Renyu Luo$^{1}$}
\author{Tian Huang$^{1}$}

\affiliation{%
 $^{1}$School of Physical Science and Technology, Xinjiang University, Urumqi, 830046, China\\
}%

\date{September 28, 2026}

\begin{abstract}
Double white dwarfs (DWDs) in the Laser Interferometer Space Antenna (LISA) band are major Galactic gravitational-wave sources, yet the standard isolated-binary picture is incomplete: a substantial fraction of tight DWD progenitors may reside in hierarchical triples.
Triple evolution can therefore alter both the production rate and the parameter distributions of LISA DWDs.
Using the Multiple Star Evolution (MSE) code, we study how the adiabatic mass-loss model (Ge model) and the SCATTER formalism for triple common-envelope (TCE) evolution affect triple-channel LISA DWD formation.
Our main model (Model 1) evolves $10^5$ triples adopting both ingredients. Three additional triple control models, each with $10^4$ triples, vary the mass-transfer stability criterion to the original MSE polytropic model, remove the SCATTER formalism, or adopt low metallicity ($Z=0.001$).
We also evolve $10^6$ isolated binaries with the same binary-evolution framework to provide a binary-channel reference.
Systems classified as dynamically unstable under the polytropic model can instead undergo stable Roche-lobe overflow (RLOF) under the Ge model.
This sustained mass transfer modifies the donor envelope, reduces its binding energy, and helps the subsequent common-envelope (CE) phase form a close DWD rather than a merger.
The Ge model increases the triple-channel LISA DWD yield by about $60\%$ relative to the polytropic model, whereas the SCATTER formalism changes the integrated yield by only about $3\%$.
After residence-time weighting, triples contribute about $1.3 \times 10^7$ LISA-band DWDs, including $1.4 \times 10^4$ individually resolvable systems. In the combined triple-plus-binary normalization, the binary channel contributes $3.1 \times 10^6$ LISA-band DWDs and $4.7 \times 10^3$ individually resolvable systems. These values are the binary share of the combined population, not the binary-only Galactic estimate.
The main triple model has a flatter, higher-$q$ mass-ratio distribution than the polytropic-model triple populations, is dominated by He-CO and He-He binaries, and produces rare eccentric systems only when a tertiary star remains bound.
\end{abstract}

\maketitle

\section{Introduction}

Gravitational waves (GWs) provide direct probes of compact binaries. The ground-based LIGO/Virgo detectors have already detected many black-hole and neutron-star mergers \citep{2017PhRvL.119p1101A}. The Laser Interferometer Space Antenna (LISA) will extend GW astronomy to the millihertz (mHz) band. It consists of three spacecraft in an approximately equilateral constellation with arm lengths of $2.5 \times 10^6\,{\rm km}$, follows the Earth around the Sun, and is planned for a 4 yr nominal mission with launch in the mid-2030s \citep{2024arXiv240207571C}. LISA is designed to detect sources with frequencies of approximately $10^{-4}-10^{-1}\,{\rm Hz}$, including the large Galactic population of compact binaries.
Double white dwarfs (DWDs) are expected to dominate the Galactic LISA source population \citep{2001A&A...365..491N, 2010ApJ...717.1006R, 2011CQGra..28i4019M}. White dwarfs (WDs) are the evolutionary endpoints of most low to intermediate-mass stars. The Milky Way is therefore expected to contain a large reservoir of DWDs. Population-synthesis studies predict millions of LISA-band DWDs, including thousands to tens of thousands of individually resolvable systems \citep{2001A&A...365..491N, 2017MNRAS.470.1894K, 2019MNRAS.490.5888L, 2022MNRAS.511.5936K, 2023ApJ...945..162T, 2023A&A...669A..82L, 2024MNRAS.534.1707T}. A smaller subset may also be identified through joint GW and electromagnetic observations \citep{2005ApJ...633L..33S, 2006CQGra..23S.809S, 2017MNRAS.470.1894K, 2019MNRAS.483.5518K, 2018MNRAS.480..302K, 2020ApJ...901....4B, 2023MNRAS.522.5358F}.
Most previous LISA DWD population studies have focused on isolated binary evolution \citep{2001A&A...365..491N, 2004MNRAS.349..181N, 2010ApJ...717.1006R, 2012A&A...546A..70T, 2014LRR....17....3P, 2019MNRAS.490.5888L}. However, triple-star evolution can also contribute to LISA DWD formation \citep{2020A&A...640A..16T, 2024ApJ...969...68H, 2025ApJ...978...47S}. The relevance of this channel is supported by multiplicity surveys and reviews: about $10\%$ of solar-type stars and about $40\%$ of B-type stars with initial masses of $5-9\,M_\odot$ are found in triple systems \citep{2008MNRAS.389..869E, 2010ApJS..190....1R, 2013ARA&A..51..269D, 2014AJ....147...86T, 2014AJ....147...87T, 2017ApJS..230...15M}. Since LISA DWD progenitors often require mass transfer \citep{1984ApJ...277..355W, 1998MNRAS.296.1019H, 2012Sci...337..444S, 2013ApJ...764..166D, 2023A&A...669A..82L}, triples can modify the formation pathway through three-body dynamics, Roche-lobe overflow (RLOF), and common-envelope (CE) evolution.

Triple evolution differs from isolated binary evolution both dynamically and in orbital architecture. Observational surveys indicate that close spectroscopic binaries are strongly associated with tertiary stars: \citet{2006A&A...450..681T} found that the fraction of solar-type spectroscopic binaries with additional companions rises to about $96\%$ for $P_{\rm in}<3\,{\rm d}$, compared with about $36\%$ for $P_{\rm in}>12\,{\rm d}$. Together with the period and multiplicity statistics summarized by \citet{2017ApJS..230...15M}, this suggests that close inner binaries in triples are either formed preferentially in compact hierarchical configurations or are further hardened by angular-momentum exchange and tertiary-induced dissipation. Triple-origin LISA DWD progenitors therefore cannot be modeled as isolated binaries whose outer companions only provide a negligible perturbation.

A bound tertiary star can also provide observational diagnostics, for example through Doppler modulation of the GW signal \citep{2008ApJ...677L..55S, 2018CQGra..35j5011R, 2018PhRvD..98f4012R}. More importantly for formation, in inclined hierarchical triples the tertiary star can drive von Zeipel-Lidov-Kozai (ZLK) oscillations, in which the inner eccentricity cycles while the secular semi-major axis remains nearly constant \citep{1910AN....183..345V, 1962P&SS....9..719L, 1962AJ.....67..591K, 2016ARA&A..54..441N}.
The relevant separation for the onset of mass transfer is then controlled not only by $a_{\rm in}$ but also by the pericenter distance, $r_{\rm p}=a_{\rm in}(1-e_{\rm in})$. ZLK excitation can drive $r_{\rm p}$ to values much smaller than the semi-major axis itself, while tidal dissipation near pericenter removes orbital energy and can circularize the inner binary at a much smaller separation \citep{2007ApJ...669.1298F, 2020A&A...640A..16T}.
These coupled processes can bring the inner binary to RLOF earlier than in isolation. At this stage, the mass ratio is often close to the critical value separating stable and unstable mass transfer, making the outcome particularly sensitive to the adopted stability criterion.

The mass-transfer stability criterion is therefore a key uncertainty. In binary evolution, the adiabatic mass-loss model (Ge model) of \citet{2020ApJS..249....9G} gives higher critical mass ratios for many giant donors than the original MSE polytropic model \citep{2010ApJ...717..724G, 2015ApJ...812...40G, 2020ApJS..249....9G}. This can stabilize RLOF and has been used to interpret the black-hole mass peak observed by LIGO, the formation of high-mass black holes, and double-neutron-star mergers \citep{2019MNRAS.490.3740N, 2021A&A...651A.100O, 2021ApJ...920...81S}. For binary-channel LISA DWDs, however, \citet{2023A&A...669A..82L} found fewer individually detectable systems with the Ge model than with the polytropic model in their fiducial CE model.
In triples, however, the dominant LISA DWD pathway is often tertiary-induced mass transfer \citep{2025A&A...704A.156R}. Because ZLK excitation and tides can make RLOF occur earlier than in an isolated binary, the Ge model may affect triples differently from binaries. Stable mass transfer can increase the mass of the inner secondary, modify the donor envelope, and alter the binding energy available in later CE evolution. These changes may affect both the triple-channel LISA DWD yield and the rate of Type Ia supernovae from DWD mergers \citep{2012ApJ...749L..11B, 2011MNRAS.417..408R, 2018MNRAS.476.2584M}.

In this work, we use the MSE code to systematically quantify the isolated effects of (i) the Ge model and (ii) the SCATTER TCE formalism on triple-channel LISA DWD formation. We combine MSE simulations with population-synthesis weighting to estimate observable and resolvable Milky Way populations and to examine the physical processes that drive differences between models. Section~\ref{sec:model} describes the simulation setup and normalization method. Section~\ref{sec:results} presents the model dependence, case studies, and predicted LISA populations. Section~\ref{sec:conclusion} summarizes the main conclusions.
\section{Model}
\label{sec:model}

All triple simulations use the open-source MSE code \citep{2021MNRAS.502.4479H}.
Each system is evolved from the zero-age main sequence (ZAMS) to a maximum age of $13.7\,{\rm Gyr}$.
Unless otherwise stated, the initial metallicity is $Z=0.02$.
The main triple model (Model 1) contains $10^5$ triples and includes both the Ge model and the SCATTER formalism for triple common-envelope (TCE) evolution.
We also simulate three additional triple models, each with $10^4$ systems: the polytropic model without the SCATTER formalism (Model 2), the Ge model without the SCATTER formalism (Model 3), and the Ge model with the SCATTER formalism at low metallicity, $Z=0.001$ (Model 4).
As a binary-channel reference, we run $10^6$ binaries with the Binary Star Evolution (BSE) code using either the Ge model (Model 5) or the polytropic model (Model 6).
The population-synthesis procedure described below is then used to estimate the Milky Way contribution of each channel.
For clarity, the input physics and sample size of each model are summarized in Table~\ref{tab:model_definitions}.

\begin{table*}[tbp]
\caption{Definitions of the population-synthesis models. Columns list the model label, evolutionary channel, simulated sample size $N_{\rm sim}$, metallicity $Z$, mass-transfer stability criterion, and TCE formalism. The TCE formalism applies only to triple systems.}
\label{tab:model_definitions}
\begin{ruledtabular}
\begin{tabular}{lccccc}
Model & Channel & $N_{\rm sim}$ & $Z$ & Mass-transfer stability criterion & TCE formalism \\
Model 1 & Triple & $10^5$ & 0.02 & Ge model & SCATTER \\
Model 2 & Triple & $10^4$ & 0.02 & Polytropic model & Original MSE \\
Model 3 & Triple & $10^4$ & 0.02 & Ge model & Original MSE \\
Model 4 & Triple & $10^4$ & 0.001 & Ge model & SCATTER \\
Model 5 & Binary & $10^6$ & 0.02 & Ge model & -- \\
Model 6 & Binary & $10^6$ & 0.02 & Polytropic model & -- \\
\end{tabular}
\end{ruledtabular}
\end{table*}

We denote the inner primary, inner secondary, and tertiary star by $M_1$, $M_2$, and $M_3$.
We draw $M_1$ from the initial mass function of \citet{2001MNRAS.322..231K}, restricted to $1-8\,M_\odot$, and sample $M_2$ using the mass-ratio prescription of \citet{2017ApJS..230...15M}.
The mutual inclination is sampled uniformly in $\cos i$.
We also use the distributions of \citet{2017ApJS..230...15M} for the inner and outer eccentricities.
When sampling $M_3$, the inner binary is treated as a single object of mass $M_1+M_2$ located at the inner-binary center of mass.
Following \citet{2023ApJ...950....9R} and the Multiple Stellar Catalog (MSC) prescription \citep{2018ApJS..235....6T}, we allow some tertiary stars to be more massive than the inner binary.
Systems that are already in RLOF \citep{1983ApJ...268..368E} or dynamically unstable \citep{2001MNRAS.321..398M, 2022MNRAS.516.4146V} at initialization are rejected and resampled.

\begin{figure*}[tbp]
\centering
\includegraphics[width=0.95\textwidth]{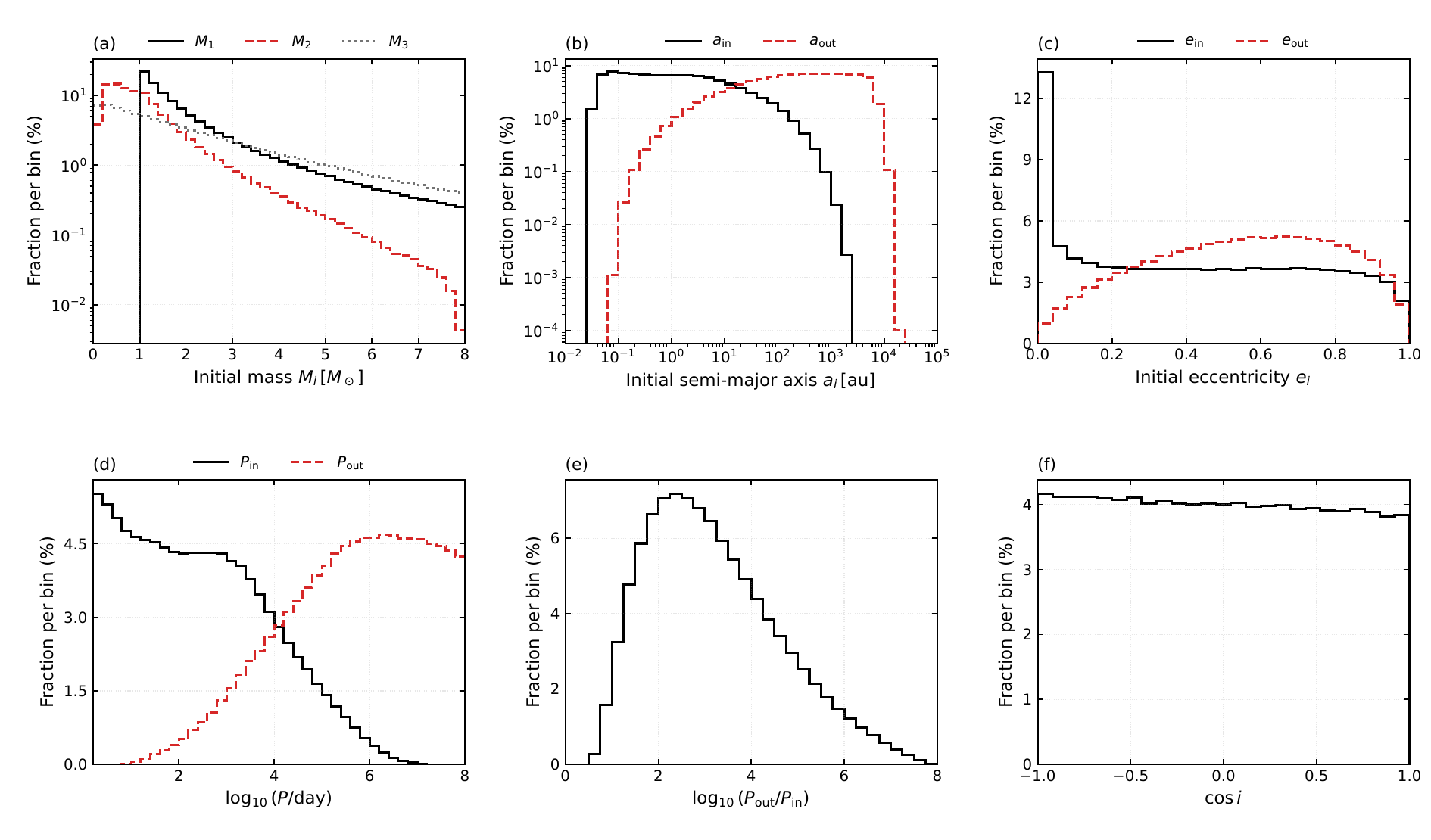}
\caption{Initial parameter distributions. The horizontal axes show (a) stellar mass, (b) semi-major axis, (c) eccentricity, (d) orbital period, (e) the outer-to-inner period ratio, and (f) the cosine of the mutual inclination. The vertical axes give the percentage of all sampled systems in each bin. In panel (a), black solid, red dashed, and gray dotted lines represent the inner primary, inner secondary, and tertiary, respectively. In panels (b)--(d), black solid and red dashed lines represent the inner and outer orbits, respectively. Black solid lines in panels (e) and (f) show the period-ratio and mutual-inclination distributions.}
\label{fig:initial_distributions}
\end{figure*}

Figure~\ref{fig:initial_distributions} shows the initial parameter distributions after applying the adopted sampling and stability-selection procedures. The mass distributions favor low-mass stars, while outer orbits generally have larger separations and longer periods than inner orbits. The period-ratio distribution follows from the joint sampling of the two periods and the stability selection. Mutual orientations are initially isotropic, with $\cos i$ sampled uniformly over $[-1,1]$, while the inclination-dependent stability selection introduces a small departure from uniformity.

\subsection{Model and Parameter Settings}

\subsubsection{MSE Program}

MSE is a C/C++ code with a Python interface that evolves hierarchical multiple systems.
For triples that satisfy the hierarchical stability criterion \citep{2001MNRAS.321..398M, 2022MNRAS.516.4146V}, MSE uses the secular approximation to follow long-term orbital evolution \citep{2016MNRAS.459.2827H, 2018MNRAS.476.4139H, 2020MNRAS.494.5492H}. Systems that become dynamically unstable are passed to direct N-body integration with NSTAR \citep{2020MNRAS.492.4131R}.
We use the default MSE treatment of single-star evolution, based on the fitting formulae of \citet{2000MNRAS.315..543H}, and the \citet{2002MNRAS.329..897H} prescriptions for wind mass loss, tides, collisions, and mergers. Similar rapid population-synthesis ingredients are widely used in compact-binary studies \citep{2008ApJS..174..223B, 2017PASA...34...58E, 2020ApJ...898...71B, 2020RAA....20..161H, 2022ApJS..258...34R}.
\subsubsection{Common Envelope Evolution}

Most LISA DWD progenitors experience one or two mass-transfer events.
After RLOF begins, the mass transfer can remain stable or become dynamically unstable depending on the mass ratio, accretion rate, and evolutionary stage of the donor.
Unstable mass transfer leads to CE evolution.
If the orbital energy released during inspiral is sufficient to eject the envelope binding energy, the two stellar cores survive as a closer binary. Otherwise, the system merges.
We adopt the standard $\alpha-\lambda$ energy formalism for CE evolution \citep{1976IAUS...73...75P, 1984ApJ...277..355W, 1988ApJ...329..764L, 1990ApJ...358..189D, 2000A&A...360.1043D, 2013A&ARv..21...59I, 2014A&A...563A..83C} as implemented in MSE, with the inner-orbit CE efficiency set to $\alpha_{\rm CE}=1$ for all models.

\subsubsection{SCATTER Common-Envelope Formalism for Triples}

In the original MSE treatment, outer-orbit RLOF is handled by treating the inner binary as a single object of mass $M_1+M_2$ in a binary with the tertiary star.
If the outer mass ratio $q_{\mathrm{out}}$ exceeds the critical value, the system enters TCE.
MSE then applies an $\alpha-\lambda$-like treatment to the outer binary but leaves the inner orbit unchanged.
This approximation neglects the dynamical response of the inner orbit to the shared envelope.
We therefore implement the SCATTER common-envelope formalism of \citet{2026MNRAS.547ag192D}.
The SCATTER formalism assumes that star-envelope interactions mediate angular-momentum exchange during CE evolution and maps the pre-CE state to the post-CE orbital separation through analytic formulae.
In hierarchical triples, this allows the inner orbit to shrink and, in some cases, the inner binary to merge during TCE.
When the tertiary star fills its Roche lobe and triggers TCE, the final outer-orbit separation $a_{\mathrm{out}}(f)$ relative to the initial value $a_{\mathrm{out}}(0)$ is:
\begin{equation}
\begin{split}
\frac{a_{\mathrm{out}}(f)}{a_{\mathrm{out}}(0)} = & \left( \frac{(M_1 + M_2) + M_3^{\mathrm{c}}}{(M_1 + M_2) + (M_3^{\mathrm{c}} + M_3^{\mathrm{env}})} \right) \\
& \times \left( \frac{M_3^{\mathrm{c}} + M_3^{\mathrm{env}}}{M_3^{\mathrm{c}}} \right)^2 \times Y ,
\end{split}
\label{eq:scatter_aout}
\end{equation}
where $M_1$, $M_2$, and $M_3$ are the masses of the inner primary, inner secondary, and tertiary star, and $M_3^{\mathrm{c}}$ and $M_3^{\mathrm{env}}$ are the tertiary core and envelope masses.
The factor $Y$ is
\begin{equation}
Y = \exp \left[ -\frac{2 M_3^{\mathrm{env}}}{(M_1 + M_2) + M_3^{\mathrm{c}}} \mathcal{F}(M_3^{\mathrm{c}}/M_{\mathrm{bin}}) \right] ,
\label{eq:scatter_y}
\end{equation}
where $\mathcal{F}$ is the angular-momentum transfer function,
\begin{equation}
\begin{split}
\mathcal{F}(q_{\mathrm{c},j}) & = \frac{\eta_{\mathrm{c}}}{q_{\mathrm{c},j}} \mathcal{Q}(q_{\mathrm{c},j}) + \frac{\eta_j}{q_{j,\mathrm{c}}} \mathcal{Q}(q_{j,\mathrm{c}}) \\
& = \mathcal{F}(q_{j,\mathrm{c}}) .
\end{split}
\label{eq:scatter_f}
\end{equation}
For the TCE process in a hierarchical triple, $q_{\mathrm{c},j}=M_3^{\mathrm{c}}/(M_1+M_2)$.
The parameter $\eta$ describes angular-momentum transfer and $\mathcal{Q}$ is the envelope mass-allocation function:
\begin{equation}
\mathcal{Q}_{\delta}(q_{\mathrm{c},j}) = \frac{f(q_{\mathrm{c},j})^\delta}{f(q_{\mathrm{c},j})^\delta + f(q_{j,\mathrm{c}})^\delta} .
\label{eq:scatter_q}
\end{equation}
\begin{equation}
\log_{10}[\eta] = -A \log_{10} \left[ \frac{M^{\mathrm{interact}}}{M_{\mathrm{tot}}(f)} \right] + B ,
\label{eq:scatter_eta}
\end{equation}
where $\delta = 3$, $M_{\mathrm{tot}}(f)$ is the total post-TCE system mass, and $M^{\mathrm{interact}}$ is the interacting envelope mass, usually taken to be $M^{\mathrm{env}}$.
The fitted constants are $A=0.95$ and $B=0.6$.
The Roche-lobe function $f(q)$ follows \citet{1983ApJ...268..368E}:
\begin{equation}
f(q) = \frac{0.49q^{2/3}}{0.6q^{2/3} + \ln(1 + q^{1/3})}.
\label{eq:eggleton}
\end{equation}
\subsection{The Adiabatic Mass-loss Model}

The critical mass ratio for dynamically unstable mass transfer determines whether, and when, a system enters CE evolution.
In standard prescriptions, stability is evaluated by comparing the donor radius response to mass loss, $\zeta_{\mathrm{ad}}$, with the Roche-lobe response, $\zeta_{\mathrm{RL}}$.
If $\zeta_{\mathrm{ad}} \le \zeta_{\mathrm{RL}}$, mass transfer proceeds on a dynamical timescale \citep{1987ApJ...318..794H, 1997A&A...327..620S, 1997MNRAS.291..732T}.
Physically, the donor expands relative to its Roche lobe and can engulf the companion, leading to CE evolution \citep{1976IAUS...73...75P}.
In the polytropic model \citep{1987ApJ...318..794H}, the giant-branch critical mass ratio $q_{\mathrm{c}}$ defined by $\zeta_{\mathrm{ad}} = \zeta_{\mathrm{RL}}$ is
\begin{equation}
q_{\mathrm{c}} = 0.362 + \frac{1}{3(1 - M_{\mathrm{c}}/M_1)} ,
\label{eq:qcrit_poly}
\end{equation}
where $M_{\mathrm{c}}$ is the core mass of the donor star, and $M_1$ is the total mass of the donor star.

The polytropic model treats the stellar gas as a fully ionized ideal gas. This approximation can make a donor expand rapidly during mass loss and therefore favors CE triggering \citep{1987ApJ...318..794H, 2003MNRAS.341..662C, 2008MNRAS.387.1416C}.
It may fail when a superadiabatic surface layer is important \citep{2011AN....332..450P, 2015MNRAS.449.4415P}. \citet{2020ApJS..249....9G} improved the adiabatic mass-loss model by using realistic stellar equations of state and showed that local thermal readjustment can suppress dynamical instability when the mass-transfer rate is below the thermal-timescale rate.
The key result for giant donors is that the critical mass ratio below which mass transfer remains stable is often significantly higher than in the polytropic model \citep{2010ApJ...717..724G, 2015ApJ...812...40G, 2020ApJS..249....9G}. This allows a wider range of RLOF episodes to proceed stably, particularly when the instantaneous mass ratio lies between the polytropic and Ge critical values.
The Ge model provides a grid of critical mass ratios for donor masses from $0.1$ to $100\,M_\odot$ and evolutionary stages from ZAMS to the tip of the asymptotic giant branch (AGB).
We implement this grid in MSE and use it as the mass-transfer stability criterion for binary and multiple-star evolution.
When the tertiary star is the donor in a hierarchical triple, we use the outer-orbit mass ratio $q_{\mathrm{out}}$ in the same stability comparison.

\subsection{Population Synthesis Method}

We select all simulated systems that form a DWD and reach the LISA band, defined here as $10^{-4}-10^{-1}\,{\rm Hz}$.
Following the population-synthesis method used in our previous work \citep{2020ApJ...890...69L, 2023RAA....23b5021Z, 2025PhRvD.111d3035W, 2025ApJ...979L..37L,2025PhRvD.111j3004L, 2025A&A...700A.147W, 2025PhRvD.112j3005L, 2026A&A...706A.105L}, we weight each target system by its residence time in the LISA band and combine these weights with the stellar birth rate and initial mass distribution.
For the stellar birth rate, we adopt the time-dependent exponential form of \citet{2001A&A...365..491N}:
\begin{equation}
\mathrm{SFR}(t) = 15 \exp(-t/\tau)\,M_\odot\,{\rm yr}^{-1} ,
\label{eq:sfr}
\end{equation}
where $\tau = 7\,{\rm Gyr}$.
This star-formation history is broadly consistent with observational constraints and Galactic chemical-evolution models \citep{1991ARA&A..29..129R, 1997A&AS..123..305V, 1999MNRAS.307..857B, 2011AJ....142..197C, 2015ApJ...806...96L}.
Because the initial mass of the inner primary is restricted to $1-8\,M_\odot$ and the raw simulation samples contain only binaries or triples rather than the full stellar population, the raw LISA selection fraction is not directly a yield per unit stellar birth mass.
We therefore normalize by the total birth mass represented by the sampled systems.
The number of LISA DWDs born per unit stellar mass is
\begin{equation}
\begin{split}
Y_{\rm LISA} =
\frac{R_{\rm t} N_{\rm LISA,t}}{M_{\rm t}}
+ \frac{R_{\rm b} N_{\rm LISA,b}}{M_{\rm b}} ,
\end{split}
\label{eq:yield_per_mass}
\end{equation}
Here the subscripts ${\rm t}$ and ${\rm b}$ denote the triple and binary simulations, respectively. $N_{\rm LISA}$ is the number of selected LISA DWDs, and $M$ is the corresponding total simulated initial mass.
The factors $R_{\rm t}$ and $R_{\rm b}$ convert the simulated mass range into the fraction of the total stellar birth mass represented by triple and binary systems.
\begin{equation}
\begin{split}
R_{\rm t} =
\frac{\sum_{\rm bins} F_{\rm t}(M_1) M_{\rm t}}
{\sum_{\rm bins}\left[F_{\rm s}(M_1)M_{\rm s}
+F_{\rm b}(M_1)M_{\rm b}
+F_{\rm t}(M_1)M_{\rm t}\right]},\\
R_{\rm b} =
\frac{\sum_{\rm bins} F_{\rm b}(M_1) M_{\rm b}}
{\sum_{\rm bins}\left[F_{\rm s}(M_1)M_{\rm s}
+F_{\rm b}(M_1)M_{\rm b}
+F_{\rm t}(M_1)M_{\rm t}\right]} .
\end{split}
\label{eq:rtriple}
\end{equation}
The functions $F_{\rm s}(M_1)$, $F_{\rm b}(M_1)$, and $F_{\rm t}(M_1)$ are the mass-dependent probabilities that a system is single, binary, or triple, respectively, taken from \citet{2017ApJS..230...15M}.
Binary and triple systems are simulated in separate samples and combined through the population normalization. Averaging the adopted multiplicity probabilities over the Kroupa IMF for $1\leq M_1/M_\odot\leq8$ gives initial system fractions of $33.9\%$ binaries, $17.8\%$ triples, and $48.3\%$ singles. These fractions describe the parent population before orbital rejection. The independently chosen binary and triple sample sizes set the sampling statistics. Their ratio is not the assumed population ratio. The birth-mass weights $R_{\rm b}$ and $R_{\rm t}$ additionally account for the masses represented by each channel.
To evaluate the denominator, we generate single-star and binary samples with the same initial-mass sampling size as the triple sample.
We assume continuous star formation over 13.7 Gyr with the time dependence in Eq.~\ref{eq:sfr}.
Because delay times and mergers remove some systems from the present-day LISA population, not every LISA DWD formed in the Milky Way remains observable today.
The total observable number is therefore
\begin{equation}
\begin{split}
N_{\mathrm{obs},\mathrm{total}} = & R_{\rm t} \sum_{i=1}^{N_{\mathrm{sim},t}} \frac{\Delta M_{\mathrm{SFR},\mathrm{obs},i}}{M_{\rm t}} \\
& + R_{\rm b} \sum_{j=1}^{N_{\mathrm{sim},b}} \frac{\Delta M_{\mathrm{SFR},\mathrm{obs},j}}{M_{\rm b}} .
\end{split}
\label{eq:nobs_total}
\end{equation}
The first term gives the triple contribution and the second gives the binary contribution.
The quantities $\Delta M_{\mathrm{SFR},\mathrm{obs},i}$ and $\Delta M_{\mathrm{SFR},\mathrm{obs},j}$ are the stellar birth masses associated with the time interval during which each triple- or binary-origin system remains observable in the LISA band, computed by integrating the star-formation history in Eq.~\ref{eq:sfr}.

Here, ``observable'' denotes a system present in the adopted LISA frequency band at the current Galactic age. It does not imply that the system can be individually resolved. We reserve ``resolvable'' for the subset satisfying the SNR criterion below.

At fixed initial distributions and evolutionary prescriptions within each channel, its in-band population scales with its consistently normalized birth-mass weight. Changing the period-ratio or mutual-inclination distribution can also change the formation efficiency through tertiary-driven interactions. The relative binary and triple contributions therefore refer to the adopted initial population. Individually resolvable counts additionally depend on the SNR selection and the resulting confusion foreground.

\subsection{Signal-to-noise Ratio and Resolvability of LISA DWDs}

To estimate the number of resolvable LISA DWDs, we calculate each simulated system's residence time in the LISA band and in the subset with sufficiently high signal-to-noise ratio $\rho$ (SNR).
For LISA DWDs, the GW frequency evolution is dominated by quadrupole radiation reaction \citep{1964PhRv..136.1224P} and depends sensitively on the chirp mass $\mathcal{M}_{\rm c}$:
\begin{equation}
\dot{f}_{\mathrm{gw}} = \frac{96}{5} \frac{(G \mathcal{M}_{\rm c})^{5/3}}{\pi c^5} (\pi f_{\mathrm{gw}})^{11/3} ,
\label{eq:fdot}
\end{equation}
where $G$ is the gravitational constant, $c$ is the speed of light, and $f_{\mathrm{gw}}$ is the GW frequency.
For circular binaries,
\begin{equation}
f_{\mathrm{gw}} = 2 f_{\mathrm{orb}} ,
\label{eq:fgw}
\end{equation}
and
\begin{equation}
\mathcal{M}_{\rm c} =
\frac{(m_1 m_2)^{3/5}}{(m_1 + m_2)^{1/5}}
= M_{\rm tot}\frac{q^{3/5}}{(1+q)^{6/5}} .
\label{eq:chirp_mass}
\end{equation}
Here $q=m_2/m_1\leq 1$ and $M_{\rm tot}=m_1+m_2$.

We count systems with signal-to-noise ratio $\rho>7$ as individually resolvable \citep{2018LISA...SRD...ESA}.
The instantaneous $\rho$ for a single source is \citep{2019CQGra..36j5011R}
\begin{equation}
\rho(f) = h(f) \sqrt{\frac{T_{\mathrm{obs}}}{S_{\mathrm{n}}(f)}} ,
\label{eq:snr}
\end{equation}
where $T_{\mathrm{obs}}=4\,{\rm yr}$ is the assumed LISA mission duration.
The instantaneous GW amplitude is
\begin{equation}
h(f) = \frac{4}{d} \left( \frac{G \mathcal{M}_{\rm c}}{c^2} \right)^{5/3} (\pi f)^{2/3} c^{-2/3} ,
\label{eq:hf}
\end{equation}
where $d$ is the source distance and $S_{\mathrm{n}}(f)$ is the total noise power spectral density.

For each simulated system, the distance $d$ is assigned by Monte Carlo sampling from an analytic Milky Way density model with disk and bulge components.
The disk density is given by \citep{2008gady.book.....B}
\begin{equation}
\rho_{\mathrm{disk}}(R, Z) \propto \exp\left( -\frac{R}{R_{\mathrm{d}}} \right) \exp\left( -\frac{|Z|}{Z_{\mathrm{d}}} \right) ,
\label{eq:disk_density}
\end{equation}
where $R_{\mathrm{d}} = 2.5\,{\rm kpc}$ and $Z_{\mathrm{d}} = 0.3\,{\rm kpc}$.
The stellar density of the bulge component is
\begin{equation}
\rho_{\mathrm{bulge}}(r) \propto \exp\left( -\frac{r^2}{2 r_{\mathrm{b}}^2} \right) ,
\label{eq:bulge_density}
\end{equation}
where $r_{\mathrm{b}} = 0.5\,{\rm kpc}$.
The LISA noise model contains instrumental noise and confusion noise.
For the instrumental component we adopt the LISA science requirements \citep{2018LISA...SRD...ESA}.
Confusion noise is produced by unresolved Galactic binaries. Although analytic fits exist \citep{2017JPhCS.840a2024C, 2019CQGra..36j5011R}, the resolvable sources themselves must be removed consistently.
We therefore use an SNR-based iterative pipeline to estimate the Galactic foreground \citep{2021PhRvD.104d3019K}, following common LISA-noise treatments \citep{2006PhRvD..73l2001T, 2007PhRvD..75d3008C, 2012ApJ...758..131N, 2025A&A...704A.156R}.
After the background noise is obtained, the total resolvable number is estimated by the same residence-time weighting:
\begin{equation}
\begin{split}
N_{\mathrm{res},\mathrm{total}} = & R_{\rm t} \sum_{i=1}^{N_{\mathrm{sim},t}} \frac{\Delta M_{\mathrm{SFR},\mathrm{res},i}}{M_{\rm t}} \\
& + R_{\rm b} \sum_{j=1}^{N_{\mathrm{sim},b}} \frac{\Delta M_{\mathrm{SFR},\mathrm{res},j}}{M_{\rm b}} .
\end{split}
\label{eq:nres_total}
\end{equation}
Here $\Delta M_{\mathrm{SFR},\mathrm{res},i}$ and $\Delta M_{\mathrm{SFR},\mathrm{res},j}$ are computed over the intervals for which the corresponding system has $\rho>7$.

We estimate the statistical errors using
\begin{equation}
\begin{aligned}
N&=\sum_i w_i,\qquad
\sigma_N=\left(\sum_i w_i^2\right)^{1/2},\\
N_{\rm eff}&=\frac{(\sum_i w_i)^2}{\sum_i w_i^2}.
\end{aligned}
\label{eq:mc_uncertainty}
\end{equation}
Where $N$ is the weighted source count, $\sigma_N$ its statistical error, and $N_{\rm eff}$ the effective sample size. The weight $w_i$ is the total contribution of independent progenitor $i$, summed over its time intervals after residence-time weighting and Galactic normalization. In Table~\ref{tab:lisa_numbers}, the errors on $N_{\rm obs}$ and $N_{\rm res}$ use the weights for their respective selections. The yield error is $\sigma_Y=(R/M)\sqrt{n}$, where $n$ is the unweighted selected-progenitor count, $R$ the channel birth-mass fraction, and $M$ the total simulated initial mass. These are sampling errors at fixed model assumptions and normalization. The error bars in Fig.~\ref{fig:wd_type_distribution} use the same weighted-Poisson prescription.

\section{Results}
\label{sec:results}

We now summarize the LISA DWD yields and parameter distributions obtained from the binary and triple simulations.
Table~\ref{tab:lisa_numbers} gives the yield per unit stellar birth mass, the normalized observable Galactic number, and the number of individually resolvable systems for each model.
The following subsections examine how the Ge model and the SCATTER formalism modify the triple-channel outcome.

\subsection{Model Dependence}

The largest change in Table~\ref{tab:lisa_numbers} comes from the mass-transfer stability criterion.
In the triple channel, the Ge model without SCATTER (Model 3) produces a LISA DWD yield more than $60\%$ higher than the polytropic model (Model 2).
The SCATTER formalism has a smaller net effect: the main model (Model 1) is about $3\%$ higher than Model 3, because fewer than $10\%$ of the LISA DWDs in our samples pass through the TCE channel.
The low-metallicity model (Model 4) gives a slightly higher yield and a larger observable and resolvable population after residence-time weighting.
In the binary channel, by contrast, our Model 5/Model 6 comparison shows that the Ge model reduces the LISA DWD yield by about $30\%$ relative to the polytropic model, in the same direction as the change in individually detectable sources found by \citet{2023A&A...669A..82L} for their fiducial CE model.

\begin{table*}[tbp]
\centering
\caption{Estimated Galactic numbers of LISA DWDs. $Y_{\rm LISA}$ is the yield per unit stellar birth mass. The third column gives the residence-time-weighted Galactic number of observable DWDs in the LISA band, and the fourth column gives the subset that is individually resolvable with $\rho>7$.}
\label{tab:lisa_numbers}
\begin{ruledtabular}
\begin{tabular}{lccc}
Category & $Y_{\rm LISA}$ & $N_{\rm obs}$ & $N_{\rm res}$ \\
 & ($10^{-4}\,M_\odot^{-1}$) & ($10^6$) & ($10^3$) \\
Triple (Model 1) & $31.0\pm0.5$ & $13.2\pm0.4$ & $14.0\pm0.5$ \\
Triple (Model 2) & $19.0\pm1.3$ & $7.9\pm0.4$ & $9.6\pm0.6$ \\
Triple (Model 3) & $29.0\pm1.6$ & $12.7\pm1.0$ & $12.6\pm0.5$ \\
Triple (Model 4) & $40.0\pm2.0$ & $17.2\pm0.7$ & $24.4\pm0.6$ \\
Binary (Model 5) & $6.2\pm0.1$ & $3.1\pm0.1$ & $4.7\pm0.1$ \\
Binary (Model 6) & $9.4\pm0.1$ & $4.9\pm0.1$ & $8.1\pm0.1$ \\
Binary only & $23.0\pm0.4$ & $10.1\pm0.4$ & $11.0\pm0.2$ \\
Triple + Binary & $37.0\pm0.5$ & $16.3\pm0.4$ & $18.7\pm0.5$ \\
\end{tabular}
\end{ruledtabular}
\end{table*}

\subsubsection{Impact of the Ge Model on the LISA Production Rate}
\begin{figure*}[htbp]
\centering
\includegraphics[width=0.95\textwidth]{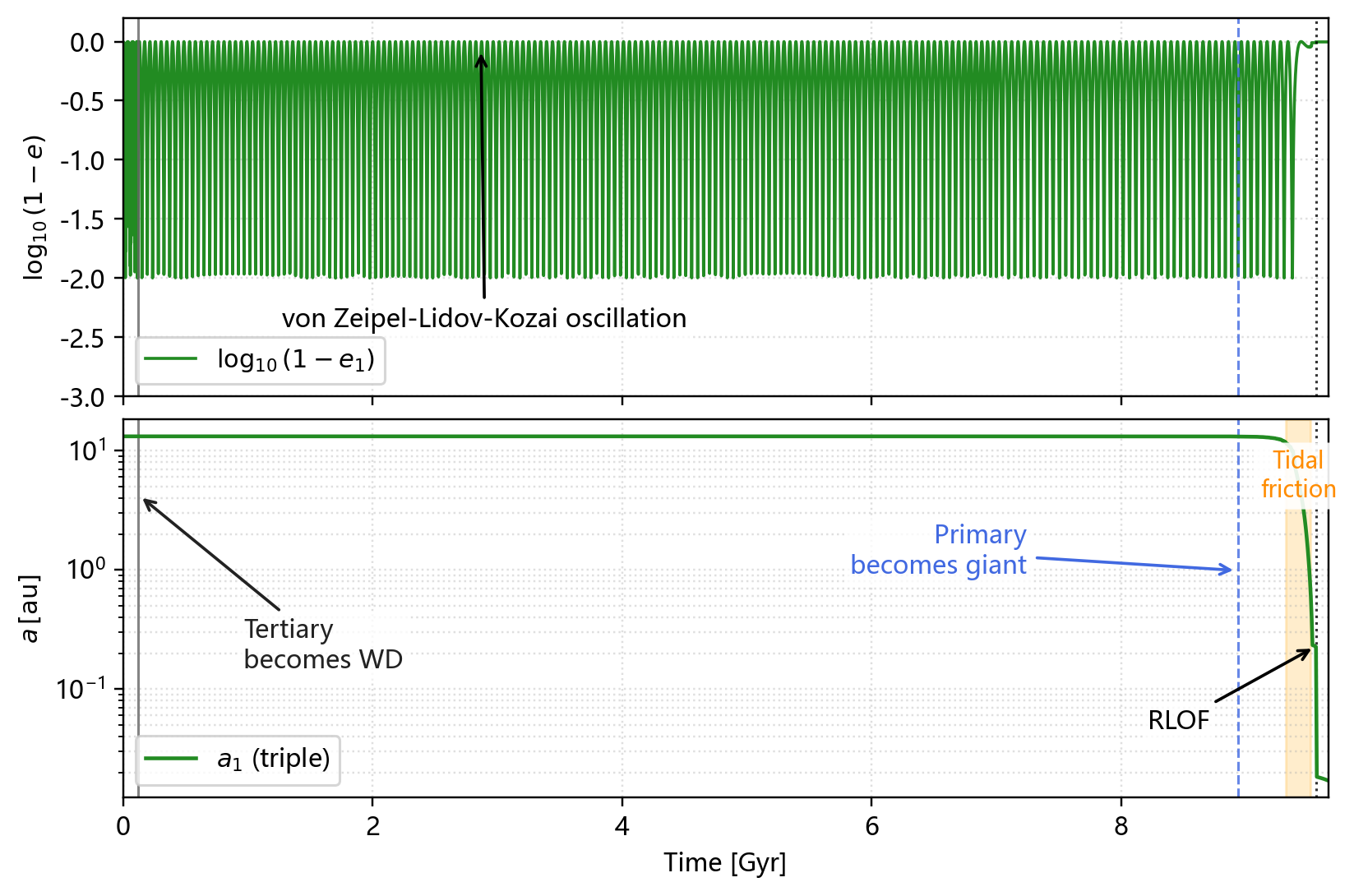}
\caption{Pre-RLOF evolution of the illustrative triple. The green solid curves show inner eccentricities $e_1$ in the upper panel and the inner semi-major axis $a_1$ in the lower panel. The gray line/black arrow marks the tertiary star becoming a WD, the blue dashed line/blue arrow marks the inner primary becoming a giant, the orange shaded band marks tidal friction, and the black RLOF arrow marks the onset of Roche-lobe overflow.}
\label{fig:pre_rlof_zlk}
\end{figure*}
The response of the LISA DWD yield to the Ge model is channel dependent. As a binary-channel reference, our Model 5/Model 6 comparison gives a reduction of about $30\%$ in the LISA DWD yield when the Ge model replaces the polytropic model. This trend agrees qualitatively with the smaller number of individually detectable sources in the fiducial Ge model of \citet{2023A&A...669A..82L}. In isolated binaries, this reduction occurs because systems that would enter an early CE under the polytropic model instead undergo stable RLOF, leaving many post-mass-transfer binaries at separations too large to reach the LISA band within a Hubble time.
The triple channel differs because the inner binary is already biased toward, or dynamically driven into, a compact configuration. Observationally, very close binaries are frequently members of higher-order systems \citep{2006A&A...450..681T, 2017ApJS..230...15M}, and dynamically the tertiary star can excite ZLK oscillations and tidal shrinkage before the initial RLOF episode \citep{2007ApJ...669.1298F, 2020A&A...640A..16T}. Thus, triples do not primarily lack a mechanism for orbital contraction. Instead, many potential LISA progenitors are removed from the LISA-forming population when the initial donor interaction becomes unstable and leads to a premature merger.

The Ge model changes this initial primary-donor interaction. For many giant donors, the Ge model gives a larger critical mass ratio than the polytropic model \citep{2010ApJ...717..724G, 2015ApJ...812...40G, 2020ApJS..249....9G}. When the instantaneous mass ratio satisfies $q_{\rm crit,poly}<q<q_{\rm crit,Ge}$, the same RLOF episode is classified as unstable in the polytropic model but stable in the Ge model.
Among the triple-channel pathways identified by \citet{2025A&A...704A.156R}, tertiary-induced mass transfer is the dominant contributor to LISA DWD formation. In this pathway, Table~\ref{tab:lisa_numbers} indicates that the Ge model increases the LISA yield by allowing primary-donor RLOF to remain stable before CE begins.
The stable phase transfers mass to the inner secondary, reduces the donor envelope, and changes the donor core and envelope structure before the subsequent CE. In the CE energy formalism, these changes affect both the envelope binding energy, $E_{\rm bind}\simeq G M_{\rm donor}M_{\rm env}/(\lambda R_{\rm donor})$, and the available orbital energy through the companion mass and the pre-CE separation.
Consequently, some systems that merge during or shortly after the primary-donor CE under the polytropic model survive the primary-donor interaction in the Ge model, undergo a later secondary-donor CE, and form close DWDs that enter the LISA band. This mechanism explains why the Ge model increases the triple-channel yield even though it decreases the binary-channel yield.
\begin{figure*}[htbp]
\centering
\includegraphics[width=0.95\textwidth]{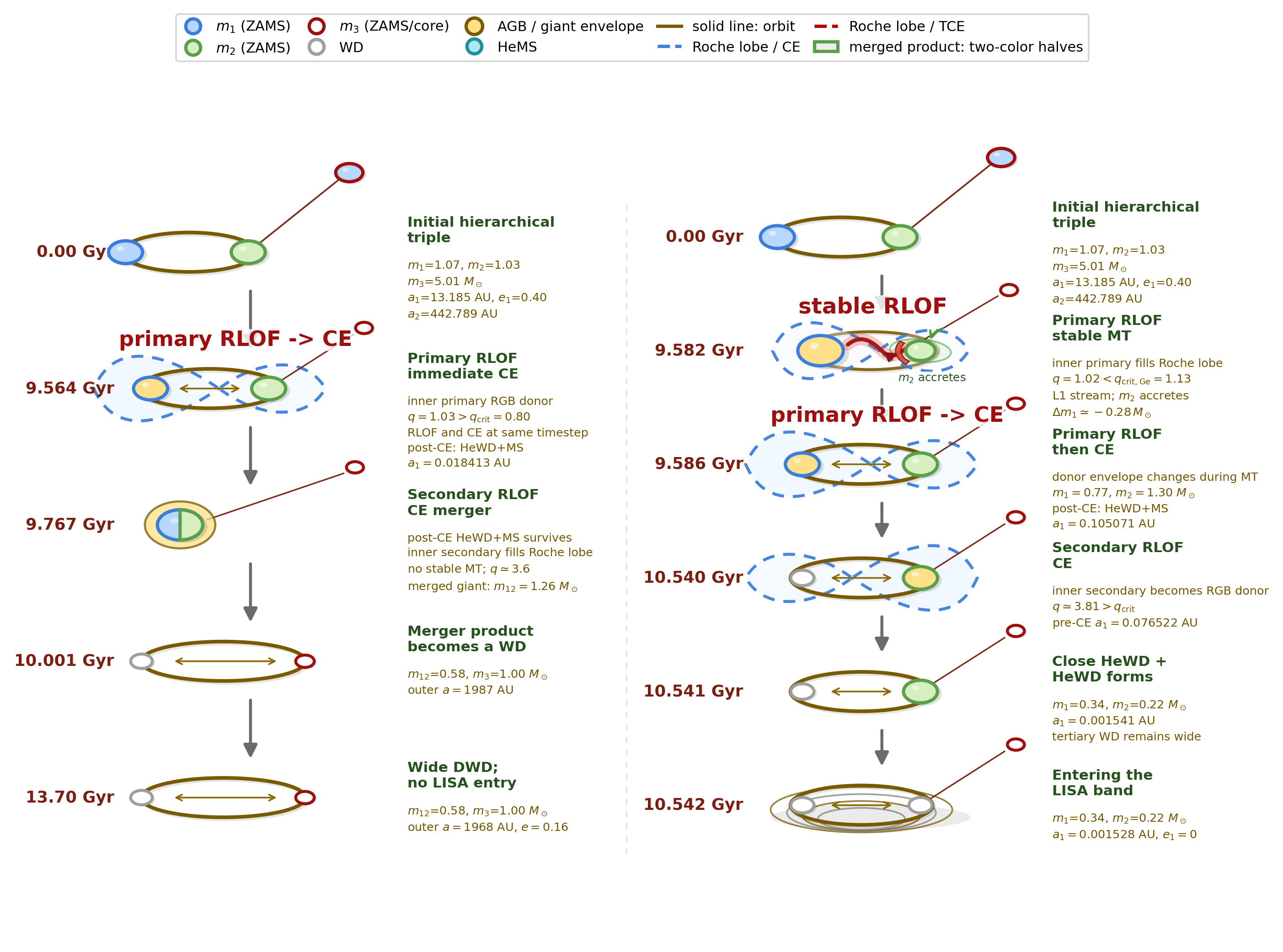}
\caption{Evolutionary outcomes for the polytropic model (left panel) and the Ge model (right panel). Blue, green, and red circles denote the inner primary $m_1$, inner secondary $m_2$, and tertiary star $m_3$/tertiary core, respectively. Gray circles are WDs, cyan circles are HeMS stars, and yellow/gold halos are giant envelopes. Brown solid ellipses are orbits, blue dashed Roche-lobe/equipotential outlines mark RLOF and CE interactions, red dashed outlines mark TCE interactions when present, gray arrows give the time sequence, and green two-color rectangles mark merger products. In the stable-RLOF step, the continuous red band shows the L1 mass-transfer stream, and the green belt marks accretion by the inner secondary.}
\label{fig:ge_case_evolution}
\end{figure*}
Figures~\ref{fig:pre_rlof_zlk} and \ref{fig:ge_case_evolution} illustrate this mechanism for one system.
The initial inner binary has masses $1.07\,M_\odot$ and $1.03\,M_\odot$, inner semi-major axis $13.2\,{\rm AU}$, and eccentricity $e=0.40$.
The tertiary star has mass $5.01\,M_\odot$, outer semi-major axis $442.8\,{\rm AU}$, outer eccentricity $e=0.25$, and mutual inclination $86^\circ$.
Figure~\ref{fig:pre_rlof_zlk} presents the pre-RLOF phase: ZLK oscillations excite the inner eccentricity, while tidal dissipation after the inner primary reaches the red giant branch (RGB) circularizes and shrinks the inner orbit to $0.2\,{\rm AU}$ \citep{1977A&A....57..383Z, 1989A&A...220..112Z, 1981A&A....99..126H, 2002MNRAS.329..897H, 2007ApJ...669.1298F}.
Figure~\ref{fig:ge_case_evolution} then compares the subsequent evolution.
In the polytropic model, RLOF begins at $9.564\,{\rm Gyr}$ when the inner primary is on the RGB.
The instantaneous mass ratio is $q=1.03$, larger than the polytropic critical value $q_{\mathrm{crit}}=0.8$, so the system enters CE.
After envelope ejection, the inner separation is only $0.02\,{\rm AU}$ and the inner primary becomes a $0.28\,M_\odot$ helium WD (HeWD).
The close orbit later causes the inner secondary to fill its Roche lobe. This secondary-donor RLOF is unstable and leads to a CE merger that produces a $1.26\,M_\odot$ giant.
Although this merger product eventually becomes a $0.58\,M_\odot$ WD and forms a DWD with the tertiary WD, the separation is too large for the system to enter the LISA band within a Hubble time.
In the Ge model, the same primary-donor RLOF episode has $q_{\mathrm{crit}}=1.13>q=1.03$, so mass transfer is initially stable.
The inner primary transfers about $0.3\,M_\odot$ through the inner L1 region to the inner secondary before CE begins.
This leaves a $0.34\,M_\odot$ helium core and a wider post-CE separation of $0.07\,{\rm AU}$.
The wider orbit avoids the premature merger, and the later CE leaves a $0.34\,M_\odot$ HeWD + $0.22\,M_\odot$ HeWD binary that enters the LISA band.

\subsubsection{The Impact of the TCE Formalism on LISA Production Rate}

\begin{figure*}[htbp]
\centering
\includegraphics[width=0.95\textwidth]{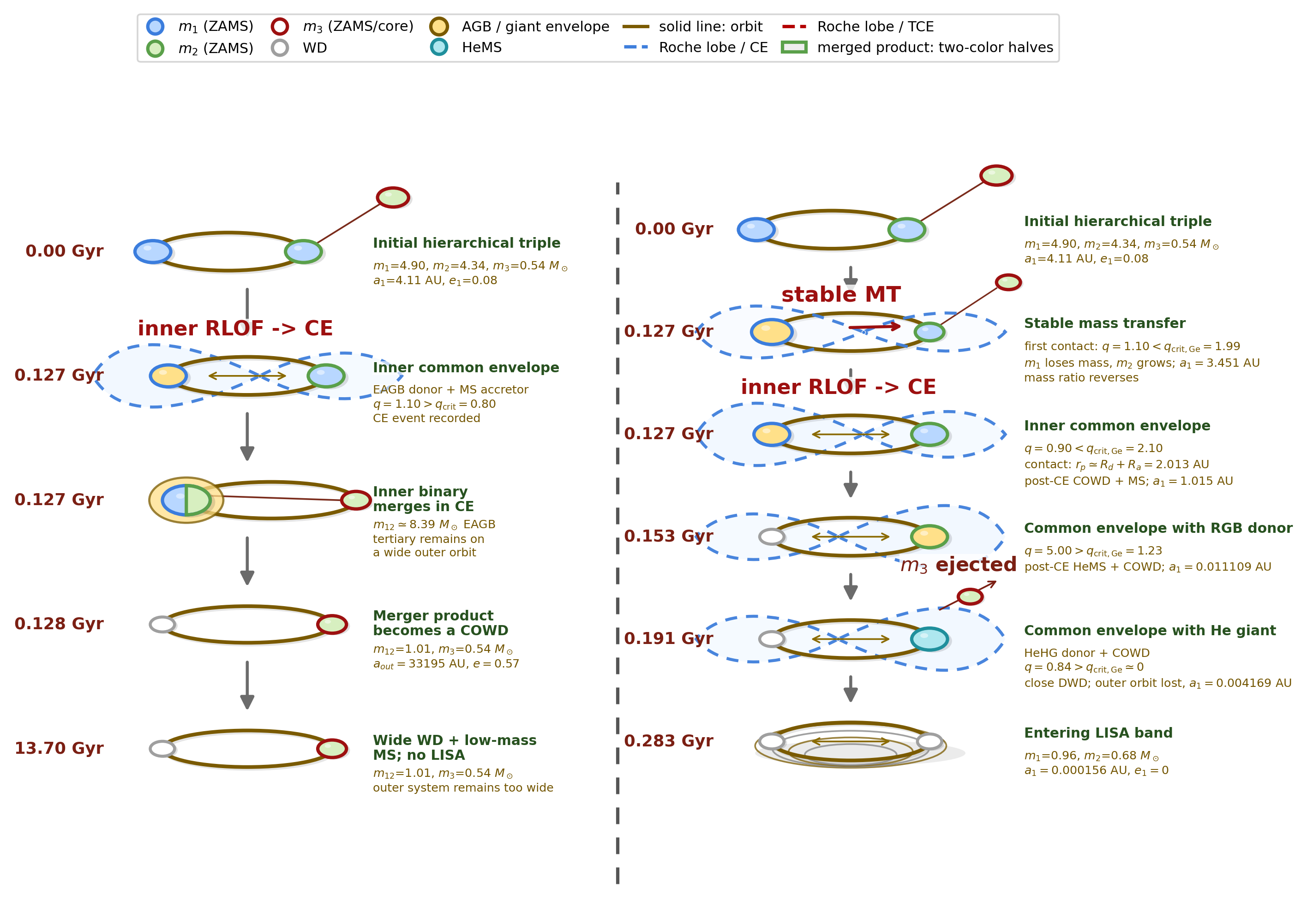}
\caption{Effect of the TCE formalism for identical initial parameters: original MSE handling of TCE (left panel) and the SCATTER formalism (right panel). Blue/green/red circles denote the inner primary $m_1$, inner secondary $m_2$, and tertiary star $m_3$/tertiary core. Gray circles are WDs, cyan circles are HeMS stars, and yellow/gold halos are giant envelopes. Brown solid ellipses are bound orbits, blue dashed Roche-lobe/equipotential outlines mark CE interactions, red dashed outlines mark tertiary RLOF and TCE interactions, gray arrows give the time sequence, and red dotted/arrowed links mark the unbound encounter and collision. Green two-color rectangles mark merger products.}
\label{fig:tce_case_evolution}
\end{figure*}

\citet{2025A&A...704A.156R} identified two TCE-related pathways to LISA DWDs. In one, TCE drives dynamical instability and inner-binary merger, after which the merger product forms a LISA DWD with the tertiary star. In the other, the tertiary star is ejected and the surviving inner binary becomes a LISA DWD.
The sensitivity to the TCE prescription comes from how the inner orbit is treated while the tertiary envelope is ejected. In the original MSE scheme, outer-orbit RLOF is reduced to a binary CE between the tertiary donor and the inner-binary center of mass, so the inner semi-major axis is kept fixed during the envelope-ejection calculation \citep{2021MNRAS.502.4479H}. If the post-TCE outer orbit becomes too compact for hierarchical stability, MSE then switches to direct N-body integration. Inner shrinkage, exchange, collision, or merger therefore occurs only after the envelope has been removed and the triple has become dynamically unstable.
The SCATTER formalism instead couples the envelope interaction to both the inner and outer orbits. It can remove angular momentum from the inner binary during TCE, allowing the inner orbit to contract or merge before the system reaches the post-envelope N-body phase. This change has a two-sided effect: inner-orbit contraction can drive systems that would otherwise avoid merger into premature inner-binary mergers, but it can also shrink otherwise too-wide post-TCE systems enough for later GW radiation to bring them into the LISA band.
Figure~\ref{fig:tce_case_evolution} gives an example with initial inner-binary masses $2.48\,M_\odot$ and $0.29\,M_\odot$, inner semi-major axis $0.2\,{\rm AU}$, inner eccentricity $0.61$, tertiary mass $7.24\,M_\odot$, outer semi-major axis $13.5\,{\rm AU}$, outer eccentricity $0.47$, and mutual inclination $35^\circ$.
In the original MSE treatment, the tertiary star reaches the thermally pulsing AGB (TPAGB), fills its Roche lobe, and triggers TCE.
After envelope ejection, the tertiary star becomes a $1.3\,M_\odot$ oxygen-neon WD (ONeWD), while the outer orbit shrinks to only $0.1\,{\rm AU}$.
Because the inner orbit has not changed, the post-TCE configuration violates the hierarchy condition and N-body integration begins.
The tertiary remnant then forms a transient, highly eccentric bound orbit with the inner secondary ($e=0.99$). The pericenter becomes smaller than the collision distance, causing the two objects to collide.
Following \citet{2002MNRAS.329..897H}, the collision product is treated as a TPAGB star with the ONeWD as its core and the low-mass main-sequence star as its envelope.
This object subsequently enters CE with the $2.48\,M_\odot$ inner primary. The low-mass envelope is ejected and a $1.31\,M_\odot$ ONeWD remains.
After several hundred Myr, the inner primary reaches the AGB and enters CE with this WD, but the post-CE orbit remains at a separation too large to reach the LISA band within a Hubble time.
With the SCATTER formalism, the same TCE episode shrinks the inner orbit and causes the inner binary to merge immediately.
The merger product remains dynamically stable with the tertiary remnant, and the exchange and collision seen in the original MSE calculation are therefore avoided.
When the merger product later reaches the RGB, it undergoes CE with the tertiary WD. The more massive envelope consumes more orbital energy, the inner primary becomes a HeWD after helium-star evolution, and the final orbit contracts to $0.01\,{\rm AU}$, allowing GW radiation to bring the DWD into the LISA band.
Overall, only about $9\%$ of our LISA DWDs experience TCE.
The SCATTER formalism can therefore change individual evolutionary histories, but its effect on the integrated yield is modest. In Table~\ref{tab:lisa_numbers}, Model 1 exceeds Model 3 by only $\sim3\%$, because most LISA DWDs are formed through non-TCE pathways and the positive and negative SCATTER outcomes partly cancel.

\subsubsection{Metallicity and Population Synthesis Results}

Models 1 and 4 isolate the metallicity dependence while keeping the Ge model and SCATTER TCE treatment fixed. Table~\ref{tab:lisa_numbers} shows that lowering the metallicity from $Z=0.02$ to $Z=0.001$ raises the triple-channel LISA DWD yield from $3.1\times10^{-3}$ to $4.0\times10^{-3}\,M_\odot^{-1}$, or by about $30\%$. After residence-time weighting, the in-band population increases from $13.2\times10^6$ to $17.2\times10^6$, while the individually resolvable population increases from $14.0\times10^3$ to $24.4\times10^3$. The resolvable fraction $N_{\rm res}/N_{\rm obs}$ also increases.

The adopted stellar-evolution model links metallicity to both wind mass loss and stellar structure \citep{2000MNRAS.315..543H}. At lower metallicity, the relevant wind losses are weaker, so the donor loses less mass before Roche-lobe overflow and retains more of its envelope at interaction.

Lower metallicity also reduces envelope opacity and generally makes giants more compact at comparable evolutionary stages \citep{2000MNRAS.315..543H}. The resulting changes in the onset and timing of Roche-lobe overflow alter the donor's core-envelope structure. A more massive envelope and a more compact donor make the common envelope more strongly bound \citep{2013A&ARv..21...59I}. Ejecting it therefore requires greater orbital-energy release and produces a tighter post-common-envelope orbit when ejection succeeds, while systems that cannot eject the envelope merge. The higher yield in Table~\ref{tab:lisa_numbers} is the net population outcome of these coupled effects. Present-day Galactic counts additionally depend on formation delays and gravitational-wave residence times, and the resolvable subset must satisfy the SNR threshold.
We do not include cosmic metallicity evolution in the population synthesis, so these numbers should be interpreted as fixed-metallicity reference yields.
The absolute contribution of low-metallicity systems may be lower if the Milky Way formed less stellar mass at low metallicity \citep{2023ApJ...945..162T}.

\begin{figure*}[htbp]
\centering
\includegraphics[width=0.95\textwidth]{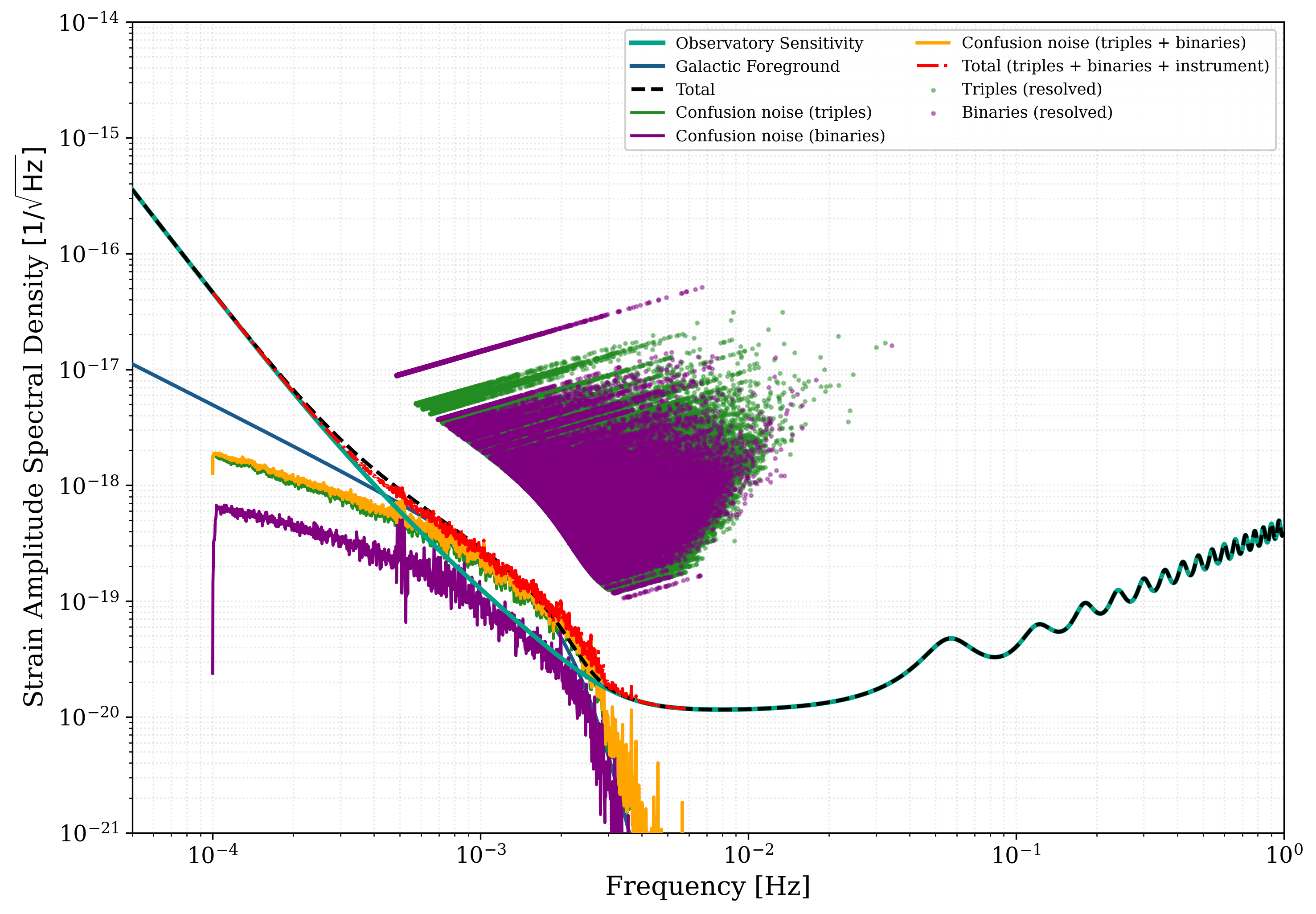}
\caption{LISA noise and resolved sources. Cyan, dark-blue, and black dashed curves show the observatory sensitivity, Galactic foreground reference, and total reference sensitivity. Green, purple, and orange solid curves show the triple, binary, and combined confusion noise. The red dash-dotted curve adds the combined confusion noise to the instrumental noise. Green and purple points are resolved triple-origin and binary-origin DWDs.}
\label{fig:lisa_noise}
\end{figure*}

Combining the yields in Table~\ref{tab:lisa_numbers} with the residence-time weighting, the main triple model predicts $1.3 \times 10^7$ observable triple-origin LISA DWDs in the Milky Way, including $1.4 \times 10^4$ resolvable systems.
In the combined triple-plus-binary normalization, the binary model contributes $3.1 \times 10^6$ observable systems and $4.7 \times 10^3$ resolvable systems. These values are the binary share of the combined population.
Together, the triple and binary channels give $1.6 \times 10^7$ observable LISA DWDs and $1.9 \times 10^4$ resolvable systems.
If triples are excluded and only the binary channel is rescaled to the full stellar population, the predicted number is $1.0 \times 10^7$ observable systems and $1.1 \times 10^4$ resolvable systems, consistent with \citet{2025A&A...704A.156R} for similar assumptions.
Figure~\ref{fig:lisa_noise} shows the resulting confusion-noise curves and resolved sources.
At low frequencies the confusion foreground is below the test-mass acceleration noise, while at high frequencies it is below the optical metrology noise.
In the range $10^{-3}$ to $10^{-2}\,{\rm Hz}$, the simulated confusion foreground is comparable to or slightly above the instrumental noise and is consistent with the level expected from LISA reference studies \citep{2017arXiv170200786A, 2024arXiv240207571C}.
The triple-origin foreground is higher than the binary-origin foreground because, in our normalization, most LISA DWDs come from the triple channel. This is consistent with \citet{2025A&A...704A.156R}.

\subsubsection{Mass Ratio and White Dwarf Type Distributions}

\begin{figure*}[tbp]
\centering
\includegraphics[width=0.90\textwidth]{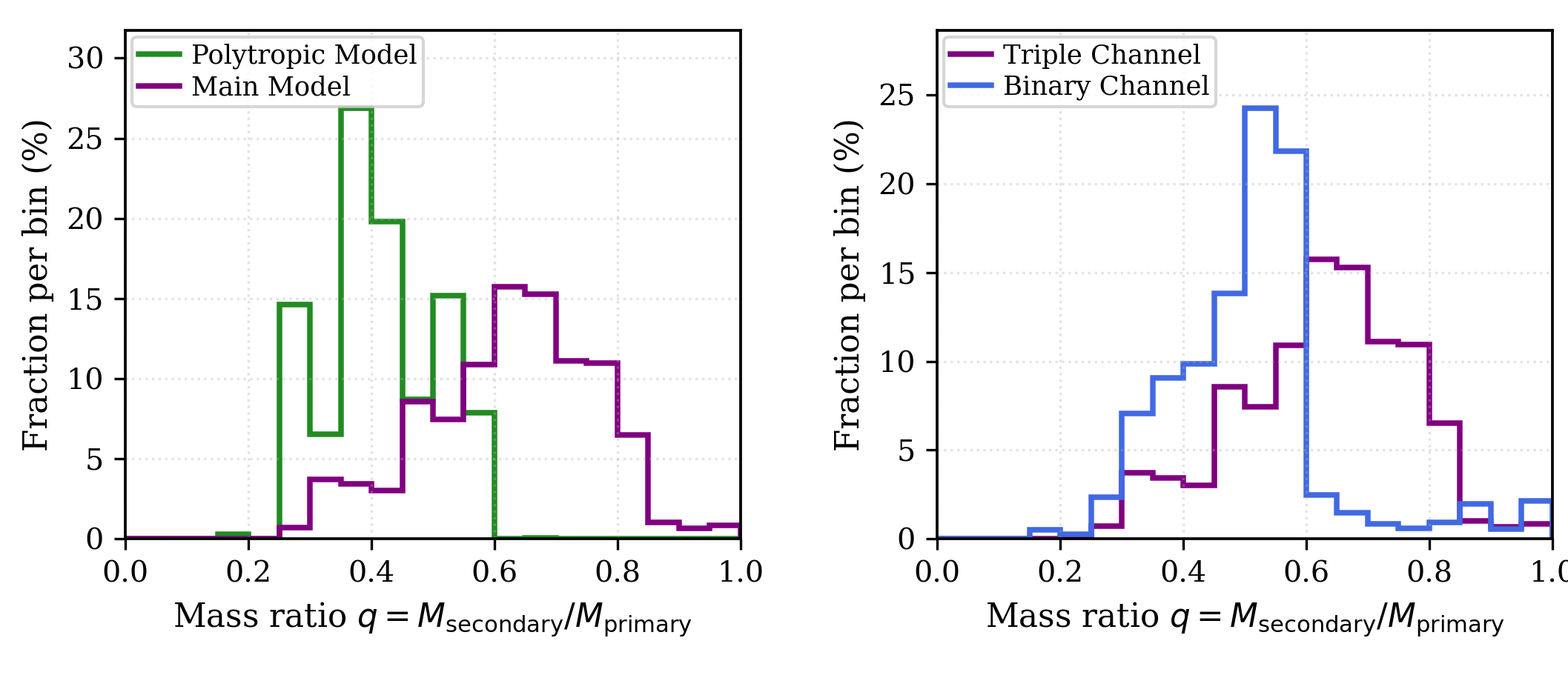}
\caption{Mass-ratio distributions. The vertical axis gives the fraction per bin. In the left panel, the green and purple solid steps show the triple-system polytropic models and the main triple model (Model 1). In the right panel, the purple and blue solid steps show the triple and binary channels.}
\label{fig:mass_ratio_distribution}
\end{figure*}

Figure~\ref{fig:mass_ratio_distribution} compares the mass-ratio distributions of the LISA DWD population.
In the mass-ratio panels, the left panel compares the main triple model with the triple-system polytropic models, and the right panel compares the main triple channel with the binary channel.
The Ge model shifts the triple-channel mass-ratio distribution toward larger values and makes it flatter than in the triple-system polytropic models.
This occurs because the Ge model allows stable RLOF over a wider range of mass ratios than the polytropic model. The resulting extended mass-transfer phase changes the component masses before CE evolution and weakens the mapping from the initial mass ratio to the final DWD mass ratio.
The right mass-ratio panel shows broadly similar channel shapes, with the triple channel slightly weighted toward larger mass ratios.

\begin{figure*}[tbp]
\centering
\includegraphics[width=0.90\textwidth]{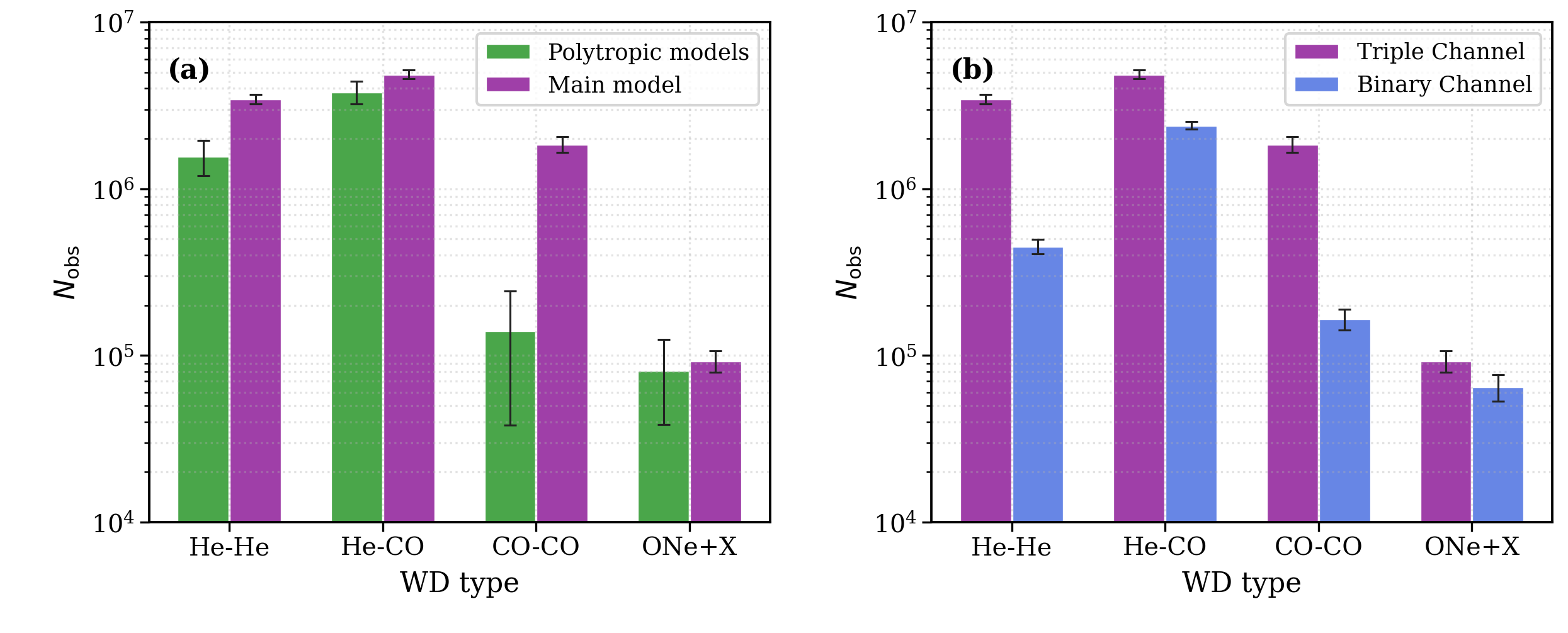}
\caption{WD-type distributions. The vertical axis gives the normalized observable Galactic number $N_{\rm obs}$. In the left panel, green and purple bars show the triple-system polytropic models and the main triple model (Model 1). In the right panel, purple and blue bars show the triple and binary channels. Error bars show weighted Poisson uncertainties, $\sqrt{\sum_i w_i^2}$, after applying the same normalization as the bar heights.}
\label{fig:wd_type_distribution}
\end{figure*}

Figure~\ref{fig:wd_type_distribution} compares the WD-type distributions.
The WD-type panels classify each DWD as He-He, He-CO, CO-CO, or ONe+X, where He, CO, and ONe WDs correspond approximately to $M<0.45\,M_\odot$, $0.45<M<1.1\,M_\odot$, and $M>1.1\,M_\odot$, respectively.
Unlike the mass-ratio panels, these panels show normalized observable Galactic numbers, $N_{\rm obs}$.
The left panel compares the main model with the triple-system polytropic models, while the right panel compares the main triple channel with the binary channel.

\subsubsection{Gravitational Wave Parameter Distributions}

\begin{figure*}[htbp]
\centering
\includegraphics[width=0.95\textwidth]{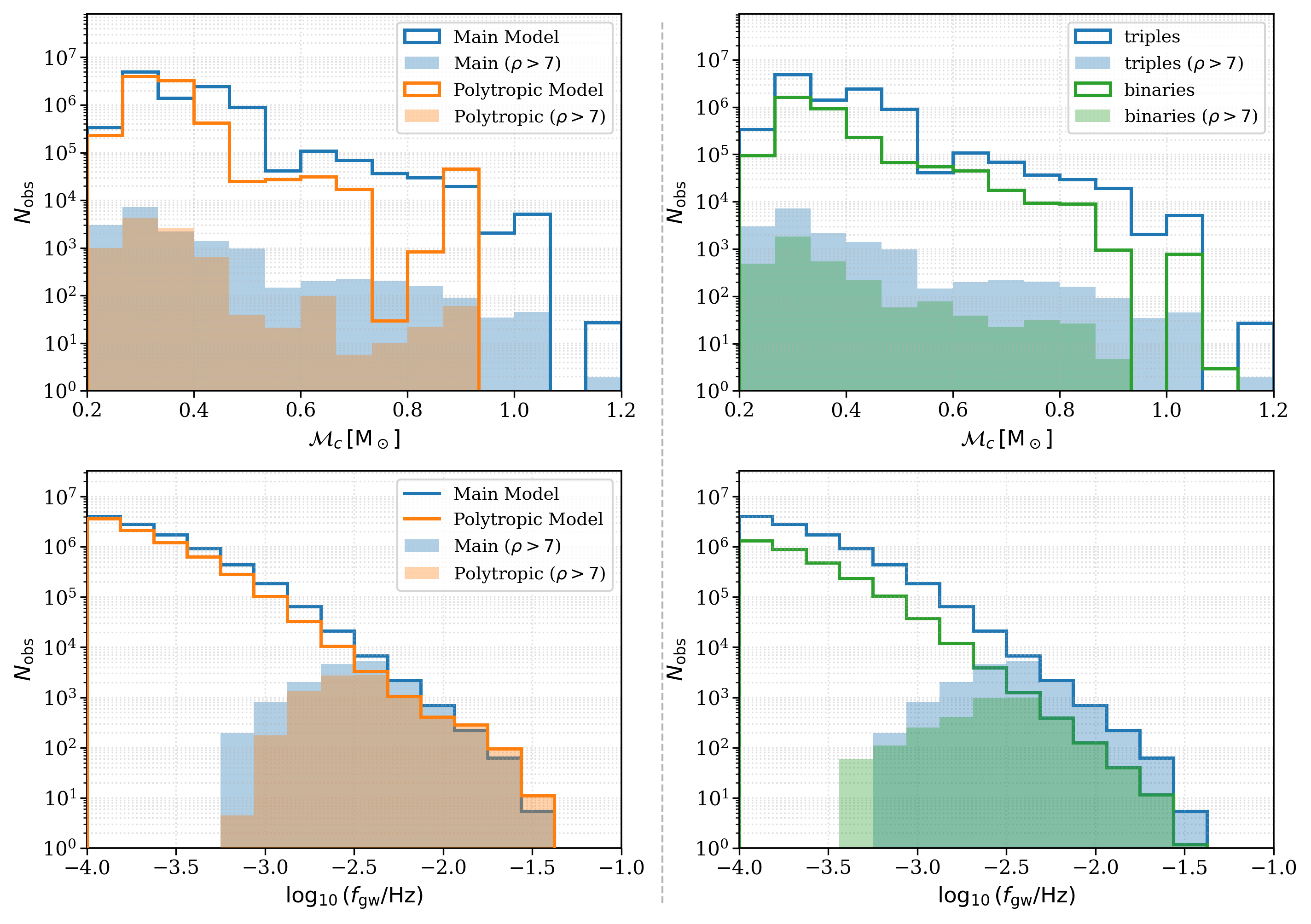}
\caption{Population-synthesis distributions of chirp mass (top) and GW frequency (bottom). In the left column, blue/orange step outlines show all systems in the main model (Model 1)/triple-system polytropic models, and blue/orange filled histograms show their $\rho>7$ subsets. In the right column, blue/green step outlines show all triple/binary-channel systems, and blue/green filled histograms show their $\rho>7$ subsets. The vertical axis gives the normalized observable Galactic number $N_{\rm obs}$.}
\label{fig:chirp_frequency}
\end{figure*}

Figure~\ref{fig:chirp_frequency} shows the chirp-mass and GW-frequency distributions for observable systems and for the $\rho>7$ subset.
The chirp-mass distribution peaks near $0.3\,M_\odot$ and is relatively broad.
In the left-column chirp-mass panels, the main model extends above $\mathcal{M}_{\rm c}=0.9\,M_\odot$, whereas the triple-system polytropic models have no weighted systems in this high-chirp-mass range.
According to Eq.~\ref{eq:chirp_mass}, $\mathcal{M}_{\rm c}=M_{\rm tot}q^{3/5}/(1+q)^{6/5}$. At fixed total mass, a larger mass ratio therefore gives a larger chirp mass.
The high-$\mathcal{M}_{\rm c}$ tail in the main model is thus consistent with Fig.~\ref{fig:mass_ratio_distribution}, where the main model is shifted toward higher mass ratios than the triple-system polytropic models.
The binary and triple channels have similar chirp-mass ranges, although the triple channel contributes more systems after normalization.
For the GW-frequency distribution, both channels show the same qualitative behavior found in previous LISA DWD studies \citep{2001A&A...365..491N, 2010ApJ...717.1006R, 2020ApJ...889...49B, 2019Natur.571..528B, 2019MNRAS.490.5888L, 2025A&A...704A.156R}.
Resolvable systems concentrate mainly at $10^{-3}-10^{-2}\,{\rm Hz}$: lower-frequency systems are limited by the foreground and instrumental noise, while high-frequency systems have shorter residence times.

\subsubsection{Eccentricity Distributions}

\begin{figure*}[tbp]
\centering
\includegraphics[width=0.96\textwidth]{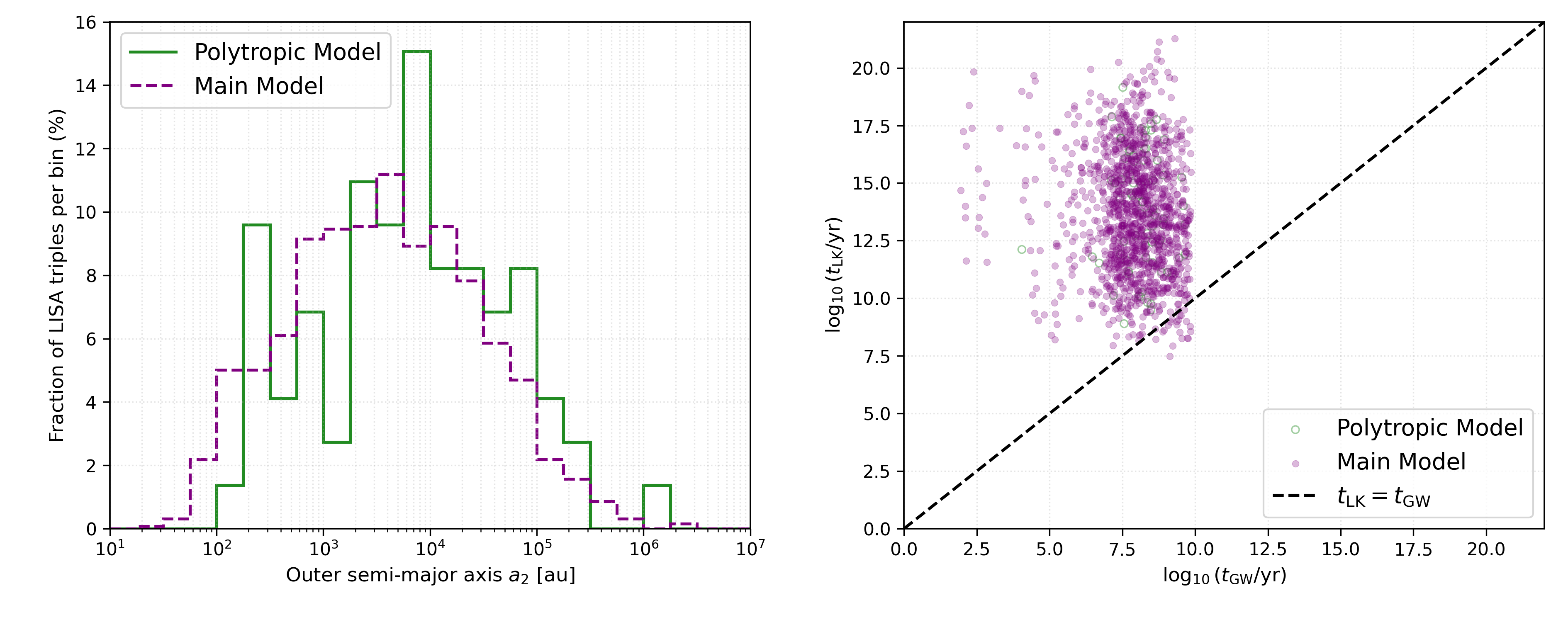}
\caption{Outer-orbit and ZLK-timescale diagnostics for LISA DWDs retaining tertiary stars. Left: green solid and purple dashed steps show the outer semi-major-axis distributions of the triple-system polytropic models and the main model (Model 1). Right: open green circles and filled purple circles show systems from the triple-system polytropic models and the main model (Model 1) in the $t_{\rm GW}$-$t_{\rm LK}$ plane. The black dashed diagonal marks $t_{\rm LK}=t_{\rm GW}$.}
\label{fig:outer_zlk}
\end{figure*}

\begin{figure*}[htbp]
\centering
\includegraphics[width=0.95\textwidth]{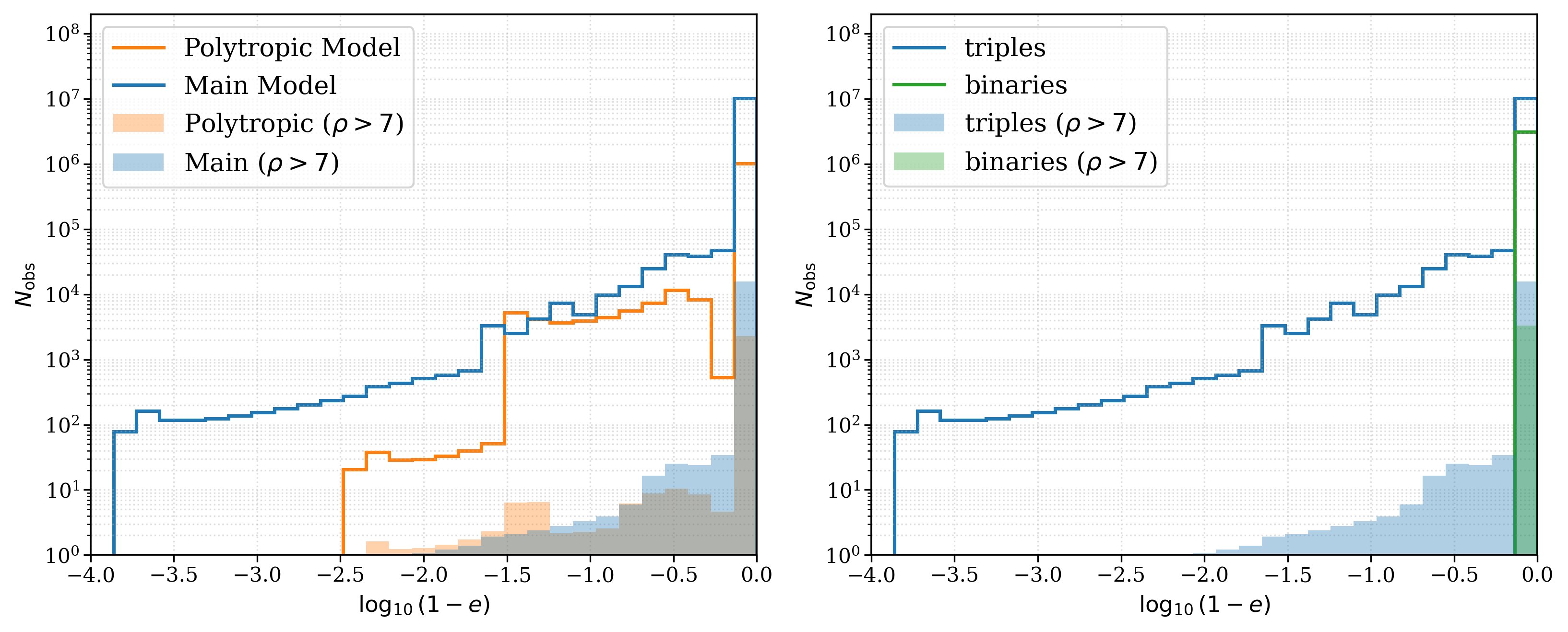}
\caption{Eccentricity distributions for the model and channel results. The vertical axis gives the normalized observable Galactic number $N_{\rm obs}$. Left: orange/blue step outlines show all systems in the triple-system polytropic models/main model (Model 1), and orange/blue filled histograms show their $\rho>7$ subsets. Right: blue/green step outlines show all triple/binary-channel systems, and blue/green filled histograms show their $\rho>7$ subsets.}
\label{fig:eccentricity_distribution}
\end{figure*}

Because LISA DWD formation usually involves CE evolution, the binary channel and most triple-channel systems are circularized.
This is especially true for the binary channel: after CE evolution, the post-CE orbit is assumed to be circularized, and no tertiary companion remains to re-excite the eccentricity through ZLK oscillations.
A small number of triple-channel systems remain eccentric because a surviving tertiary star continues to drive ZLK oscillations \citep{2025A&A...704A.156R}.
In our simulations, about half of the LISA DWDs retain an outer orbit at formation, but most tertiary stars are too distant for the ZLK timescale to be observationally relevant.
Figure~\ref{fig:outer_zlk} shows the outer semi-major-axis distribution for systems retaining tertiary stars and compares the GW inspiral timescale with the ZLK timescale.
Only systems with $t_{\rm LK}<t_{\rm GW}$ can complete at least one eccentricity-oscillation cycle before GW inspiral dominates.
The main and polytropic models have similar outer-orbit distributions, and the time-weighted eccentric LISA DWD population is only about $10^{-4}$ of the total observable population, consistent with \citet{2025A&A...704A.156R}.
Figure~\ref{fig:eccentricity_distribution} shows the corresponding eccentricity distributions for the model and channel results. We plot $\log_{10}(1-e)$ to resolve the strongly circularized population near $e=0$. In this representation, the binary-channel systems pile up at the circular boundary rather than indicating missing data.

Eccentric DWDs distribute GW power among harmonics of the orbital frequency, whereas a circular system radiates predominantly at twice the orbital frequency \citep{1963PhRv..131..435P}. If additional harmonics have sufficient SNR, their relative amplitudes can constrain the eccentricity. A measurable eccentricity would therefore provide a way to identify candidates for dynamical formation or ongoing tertiary perturbations, including the triple-origin systems found here. The outer orbit produces a Doppler shift in the GW frequency through the line-of-sight motion of the inner binary's center of mass. When the outer period greatly exceeds the observation time, a nearly constant shift is absorbed into the inferred source frequency. Any slow drift can be confused with intrinsic frequency evolution from GW emission or mass transfer if higher frequency derivatives are not measurable \citep{2018PhRvD..98f4012R}. Applying the detectability criterion of \citet{2018PhRvD..98f4012R}, Ref.~\cite{2025A&A...704A.156R} found that all surviving tertiaries in their population have outer orbits too wide to produce detectable GW imprints. Our simulations similarly produce predominantly wide outer orbits, consistent with this picture.

\subsection{Comparisons with Other Models}

Our combined triple+binary estimate gives $1.6 \times 10^7$ observable LISA DWDs in the Milky Way, of which $1.9 \times 10^4$ are individually resolvable.
The predicted number of resolvable sources is broadly consistent with previous studies, while the total LISA-band population varies by factors of a few across the literature.
Table~\ref{tab:literature_comparison} compares our estimate with representative published LISA DWD population calculations.
\begin{table}[tbp]
\footnotesize
\caption{Comparison with published LISA DWD population estimates. $N_{\rm LISA}$ and $N_{\rm res}$ denote the in-band and individually resolvable populations, respectively. Literature values retain the frequency cuts, mission duration, and detection criteria of their original studies. They have not been recalculated with our selection. The binary-only and combined populations from this work are listed separately to distinguish their normalization. Superscripts identify the references below. A dash means that value was not reported.}
\label{tab:literature_comparison}
\begin{ruledtabular}
\begin{tabular}{lccc}
 & Channel & $N_{\rm LISA}$ & $N_{\rm res}$ \\
 & & ($10^6$) & ($10^3$) \\
Nelemans (2001)$^{\mathrm{a}}$ & Binary & -- & 12.0 \\
Nelemans (2004)$^{\mathrm{b}}$ & Binary & -- & 11.0 \\
Korol (2017)$^{\mathrm{c}}$ & Binary & 26.0 & 25.0 \\
Lamberts (2019)$^{\mathrm{d}}$ & Binary & 62.0 & 12.0 \\
Li (2023), Ge$^{\mathrm{e}}$ & Binary & -- & 27.1 \\
Rajamuthukumar et al. (2025)$^{\mathrm{f}}$ & Bin.+tri. & 11.0 & 17.4 \\
This work (Binary only) & Binary & 10.1 & 11.0 \\
This work (combined) & Tri.+bin. & 16.3 & 18.7 \\
\end{tabular}
\end{ruledtabular}
\begin{flushleft}
\scriptsize
$^{\mathrm{a}}$ \citet{2001A&A...365..491N},
$^{\mathrm{b}}$ \citet{2004MNRAS.349..181N},
$^{\mathrm{c}}$ \citet{2017MNRAS.470.1894K},
$^{\mathrm{d}}$ \citet{2019MNRAS.490.5888L},
$^{\mathrm{e}}$ \citet{2023A&A...669A..82L},
$^{\mathrm{f}}$ \citet{2025A&A...704A.156R}.
The Li et al.\ row refers to their fiducial $\alpha_{\rm CE}=1$ Ge model. Their total Galactic DWD population is not a count restricted to our LISA frequency band.
\end{flushleft}
\end{table}
For binary-only studies, \citet{2017MNRAS.470.1894K} used SeBa \citep{1996A&A...309..179P} and estimated $2.6 \times 10^7$ foreground LISA DWDs, including $2.5 \times 10^4$ resolvable systems.
\citet{2019MNRAS.490.5888L} combined BSE with FIRE cosmological simulations and found about $6.2 \times 10^7$ Galactic LISA DWDs and $1.2 \times 10^4$ resolvable systems.
\citet{2001A&A...365..491N} and \citet{2004MNRAS.349..181N} reported resolvable LISA DWD populations of $1.2 \times 10^4$ and $1.1 \times 10^4$, respectively.
For their fiducial $\alpha_{\rm CE}=1$ calculation, \citet{2023A&A...669A..82L} give about $2.7\times10^4$ individually detectable DWDs with the Ge model and $3.6\times10^4$ with the polytropic model, assuming four years of LISA observations and SNR greater than 7. Their total Galactic DWD count includes systems outside the LISA band and is not used as an in-band estimate in Table~\ref{tab:literature_comparison}.
Our binary-only estimate is lower, about $1 \times 10^7$ LISA DWDs and $1.1 \times 10^4$ resolvable systems.
\citet{2025A&A...704A.156R}, which also included triples, predicted $1.1 \times 10^7$ total LISA DWDs and $1.74 \times 10^4$ resolvable systems, with the triple channel contributing most of both populations.
Differences among these studies arise from initial-parameter sampling, Galactic models, stellar birth rates, channel weighting, and stellar-evolution prescriptions.

The appropriate comparison with binary-only population studies is our ``Binary only'' calculation in Table~\ref{tab:lisa_numbers}, which assigns the stellar population consistently to single and binary systems. The binary contribution in our mixed triple+binary model represents a smaller share of the same Galactic birth mass and should not be compared directly with an entire binary-only Galaxy. Korol et al.\ use SeBa with an analytic Galactic model, whereas Lamberts et al.\ combine binary evolution with a galaxy from the FIRE simulations \citep{2017MNRAS.470.1894K,2019MNRAS.490.5888L}. Li et al.\ examine the dependence on mass-transfer stability within a binary population \citep{2023A&A...669A..82L}. These choices help explain why similar numbers of resolvable systems can accompany different total in-band populations. Our excess of triple-origin sources is assessed by comparing the two channels within the same Galactic normalization and LISA selection, while Models 5 and 6 isolate the effect of the stability criterion within the binary channel.
Within our binary-channel models, the Ge model reduces the LISA DWD yield by about $30\%$ relative to the polytropic model. This trend agrees qualitatively with the binary-population results of \citet{2023A&A...669A..82L}.
For triples, the polytropic-model triple calculation gives $7.9 \times 10^6$ LISA DWDs, consistent with the triple calculation of \citet{2025A&A...704A.156R} under similar initial sampling.
Thus the difference between our final estimate and some literature values is driven by both population normalization and the adopted mass-transfer physics.

\section{Conclusion}
\label{sec:conclusion}

We have used MSE to study how the Ge model and the SCATTER common-envelope formalism affect the formation of triple-channel LISA DWDs.
We also used BSE calculations to compare with the binary channel.
Our main conclusions are as follows.

First, the Ge model increases the triple-channel LISA DWD yield by about $60\%$ relative to the polytropic model.
This differs from the binary channel because triple progenitors are often already compact or are driven toward RLOF by tertiary-induced ZLK oscillations and tidal dissipation.
Since triple progenitors already possess efficient orbital-contraction mechanisms, the primary role of the Ge model is not to enable contraction where none exists, but to prevent premature CE mergers that would otherwise remove systems from the LISA-forming population.
By allowing the primary-donor RLOF episode to remain stable for longer, it changes the donor envelope, increases the mass of the inner secondary, and allows systems that would merge under the polytropic model to survive as close LISA-band DWDs.

Second, the SCATTER formalism for TCE can alter individual evolutionary histories but has a small net statistical effect.
By allowing inner-orbit contraction during TCE, the SCATTER formalism can shrink some systems enough to enter the LISA band, but it can also cause premature inner-binary mergers that remove potential LISA progenitors.
Because only about $9\%$ of the LISA DWDs in our samples pass through TCE, and because these positive and negative outcomes largely cancel, the integrated yield increases by only about $3\%$.

Third, after residence-time weighting, the main triple model predicts $1.3 \times 10^7$ observable triple-origin LISA DWDs in the Milky Way, including $1.4 \times 10^4$ resolvable systems.
In the combined triple-plus-binary normalization, the binary model contributes $3.1 \times 10^6$ observable systems and $4.7 \times 10^3$ resolvable systems. These values are the binary share of the combined population.
The combined estimate is therefore $1.6 \times 10^7$ observable LISA DWDs and $1.9 \times 10^4$ individually resolvable systems, with the triple channel contributing about $80\%$ of the observable population and about $75\%$ of the resolvable population under the adopted initial-population normalization. These fractions depend on the adopted initial multiplicity distribution and Galactic normalization.

Fourth, the Ge model changes the mass-ratio distribution, making the main triple model flatter and more weighted toward high mass ratios than the triple-system polytropic models.
The WD-type distributions are dominated by He-CO and He-He systems, while CO-CO systems are more prominent in the main triple model than in the triple-system polytropic models.
The GW-frequency distributions are similar between channels, and resolvable systems mainly occupy the $10^{-3}-10^{-2}\,{\rm Hz}$ range.

Finally, in our simulated populations, eccentric LISA DWDs are rare and occur only in triple-origin systems that retain tertiary stars.
Most surviving outer orbits have separations too large for rapid ZLK oscillations, and the time-weighted eccentric population is only about $10^{-4}$ of the total observable LISA DWD population.
The resulting Galactic confusion foreground is comparable to or slightly above the instrumental noise in the mHz band and is consistent with the LISA reference foreground level.

\begin{acknowledgments}
This work has been supported by the National Natural Science Foundation of China (Grants 12563007, 12541303, 12373038, 12288102, 12463011), the Natural Science Foundation of Xinjiang (Grants 2022D01D85 and 2024D01C230), the China Manned Space Program (Grant CMS-CSST-2025-A15), the Foundation of Tianshan Talents program 2024TSYCJU0001, the Major Science and Technology Program of Xinjiang Uygur Autonomous Region under grant 2022A03013-3, and the Outstanding Graduate Innovation Project of Xinjiang University with No. XJDX2025YJS039.
\end{acknowledgments}

\bibliography{PRD,Referee_B_additions}

@ARTICLE{2017PhRvL.119p1101A,
       author = {{Abbott}, B.~P. and {Abbott}, R. and {Abbott}, T.~D. and {Acernese}, F. and {Ackley}, K. and {Adams}, C. and {Adams}, T. and {Addesso}, P. and {Adhikari}, R.~X. and {Adya}, V.~B. and {Affeldt}, C. and {Afrough}, M. and {Agarwal}, B. and {Agathos}, M. and {Agatsuma}, K. and {Aggarwal}, N. and {Aguiar}, O.~D. and {Aiello}, L. and {Ain}, A. and {Ajith}, P. and {Allen}, B. and {Allen}, G. and {Allocca}, A. and {Altin}, P.~A. and {Amato}, A. and {Ananyeva}, A. and {Anderson}, S.~B. and {Anderson}, W.~G. and {Angelova}, S.~V. and {Antier}, S. and {Appert}, S. and {Arai}, K. and {Araya}, M.~C. and {Areeda}, J.~S. and {Arnaud}, N. and {Arun}, K.~G. and {Ascenzi}, S. and {Ashton}, G. and {Ast}, M. and {Aston}, S.~M. and {Astone}, P. and {Atallah}, D.~V. and {Aufmuth}, P. and {Aulbert}, C. and {AultONeal}, K. and {Austin}, C. and {Avila-Alvarez}, A. and {Babak}, S. and {Bacon}, P. and {Bader}, M.~K.~M. and {Bae}, S. and {Bailes}, M. and {Baker}, P.~T. and {Baldaccini}, F. and {Ballardin}, G. and {Ballmer}, S.~W. and {Banagiri}, S. and {Barayoga}, J.~C. and {Barclay}, S.~E. and {Barish}, B.~C. and {Barker}, D. and {Barkett}, K. and {Barone}, F. and {Barr}, B. and {Barsotti}, L. and {Barsuglia}, M. and {Barta}, D. and {Barthelmy}, S.~D. and {Bartlett}, J. and {Bartos}, I. and {Bassiri}, R. and {Basti}, A. and {Batch}, J.~C. and {Bawaj}, M. and {Bayley}, J.~C. and {Bazzan}, M. and {B{\'e}csy}, B. and {Beer}, C. and {Bejger}, M. and {Belahcene}, I. and {Bell}, A.~S. and {Berger}, B.~K. and {Bergmann}, G. and {Bernuzzi}, S. and {Bero}, J.~J. and {Berry}, C.~P.~L. and {Bersanetti}, D. and {Bertolini}, A. and {Betzwieser}, J. and {Bhagwat}, S. and {Bhandare}, R. and {Bilenko}, I.~A. and {Billingsley}, G. and {Billman}, C.~R. and {Birch}, J. and {Birney}, R. and {Birnholtz}, O. and {Biscans}, S. and {Biscoveanu}, S. and {Bisht}, A. and {Bitossi}, M. and {Biwer}, C. and {Bizouard}, M.~A. and {Blackburn}, J.~K. and {Blackman}, J. and {Blair}, C.~D. and {Blair}, D.~G. and {Blair}, R.~M. and {Bloemen}, S. and {Bock}, O. and {Bode}, N. and {Boer}, M. and {Bogaert}, G. and {Bohe}, A. and {Bondu}, F. and {Bonilla}, E. and {Bonnand}, R. and {Boom}, B.~A. and {Bork}, R. and {Boschi}, V. and {Bose}, S. and {Bossie}, K. and {Bouffanais}, Y. and {Bozzi}, A. and {Bradaschia}, C. and {Brady}, P.~R. and {Branchesi}, M. and {Brau}, J.~E. and {Briant}, T. and {Brillet}, A. and {Brinkmann}, M. and {Brisson}, V. and {Brockill}, P. and {Broida}, J.~E. and {Brooks}, A.~F. and {Brown}, D.~A. and {Brown}, D.~D. and {Brunett}, S. and {Buchanan}, C.~C. and {Buikema}, A. and {Bulik}, T. and {Bulten}, H.~J. and {Buonanno}, A. and {Buskulic}, D. and {Buy}, C. and {Byer}, R.~L. and {Cabero}, M. and {Cadonati}, L. and {Cagnoli}, G. and {Cahillane}, C. and {Calder{\'o}n Bustillo}, J. and {Callister}, T.~A. and {Calloni}, E. and {Camp}, J.~B. and {Canepa}, M. and {Canizares}, P. and {Cannon}, K.~C. and {Cao}, H. and {Cao}, J. and {Capano}, C.~D. and {Capocasa}, E. and {Carbognani}, F. and {Caride}, S. and {Carney}, M.~F. and {Carullo}, G. and {Casanueva Diaz}, J. and {Casentini}, C. and {Caudill}, S. and {Cavagli{\`a}}, M. and {Cavalier}, F. and {Cavalieri}, R. and {Cella}, G. and {Cepeda}, C.~B. and {Cerd{\'a}-Dur{\'a}n}, P. and {Cerretani}, G. and {Cesarini}, E. and {Chamberlin}, S.~J. and {Chan}, M. and {Chao}, S. and {Charlton}, P. and {Chase}, E. and {Chassande-Mottin}, E. and {Chatterjee}, D. and {Chatziioannou}, K. and {Cheeseboro}, B.~D. and {Chen}, H.~Y. and {Chen}, X. and {Chen}, Y. and {Cheng}, H.-P. and {Chia}, H. and {Chincarini}, A. and {Chiummo}, A. and {Chmiel}, T. and {Cho}, H.~S. and {Cho}, M. and {Chow}, J.~H. and {Christensen}, N. and {Chu}, Q. and {Chua}, A.~J.~K. and {Chua}, S.},
        title = "{GW170817: Observation of Gravitational Waves from a Binary Neutron Star Inspiral}",
      journal = {\prl},
         year = 2017,
        month = oct,
       volume = {119},
       number = {16},
          eid = {161101},
        pages = {161101},
          doi = {10.1103/PhysRevLett.119.161101},
archivePrefix = {arXiv},
       eprint = {1710.05832},
 primaryClass = {gr-qc},
       adsurl = {https://ui.adsabs.harvard.edu/abs/2017PhRvL.119p1101A}
}

@ARTICLE{2025ApJ...979L..37L,
       author = {{Li}, Zhuowen and {Lu}, Xizhen and {L{\"u}}, Guoliang and {Zhu}, Chunhua and {Liu}, Helei and {Yu}, Jinlong},
        title = "{A Possible Formation Scenario of the Gaia ID 3425577610762832384: Inner Binary Merger inside a Triple Common Envelope}",
      journal = {\apjl},
         year = 2025,
        month = feb,
       volume = {979},
       number = {2},
          eid = {L37},
        pages = {L37},
          doi = {10.3847/2041-8213/ada614},
archivePrefix = {arXiv},
       eprint = {2501.05139},
 primaryClass = {astro-ph.SR},
       adsurl = {https://ui.adsabs.harvard.edu/abs/2025ApJ...979L..37L}
}

@ARTICLE{2017arXiv170200786A,
       author = {{Amaro-Seoane}, Pau and {Audley}, Heather and {Babak}, Stanislav and {Baker}, John and {Barausse}, Enrico and {Bender}, Peter and {Berti}, Emanuele and {Binetruy}, Pierre and {Born}, Michael and {Bortoluzzi}, Daniele and {Camp}, Jordan and {Caprini}, Chiara and {Cardoso}, Vitor and {Colpi}, Monica and {Conklin}, John and {Cornish}, Neil and {Cutler}, Curt and {Danzmann}, Karsten and {Dolesi}, Rita and {Ferraioli}, Luigi and {Ferroni}, Valerio and {Fitzsimons}, Ewan and {Gair}, Jonathan and {Gesa Bote}, Lluis and {Giardini}, Domenico and {Gibert}, Ferran and {Grimani}, Catia and {Halloin}, Hubert and {Heinzel}, Gerhard and {Hertog}, Thomas and {Hewitson}, Martin and {Holley-Bockelmann}, Kelly and {Hollington}, Daniel and {Hueller}, Mauro and {Inchauspe}, Henri and {Jetzer}, Philippe and {Karnesis}, Nikos and {Killow}, Christian and {Klein}, Antoine and {Klipstein}, Bill and {Korsakova}, Natalia and {Larson}, Shane L and {Livas}, Jeffrey and {Lloro}, Ivan and {Man}, Nary and {Mance}, Davor and {Martino}, Joseph and {Mateos}, Ignacio and {McKenzie}, Kirk and {McWilliams}, Sean T and {Miller}, Cole and {Mueller}, Guido and {Nardini}, Germano and {Nelemans}, Gijs and {Nofrarias}, Miquel and {Petiteau}, Antoine and {Pivato}, Paolo and {Plagnol}, Eric and {Porter}, Ed and {Reiche}, Jens and {Robertson}, David and {Robertson}, Norna and {Rossi}, Elena and {Russano}, Giuliana and {Schutz}, Bernard and {Sesana}, Alberto and {Shoemaker}, David and {Slutsky}, Jacob and {Sopuerta}, Carlos F. and {Sumner}, Tim and {Tamanini}, Nicola and {Thorpe}, Ira and {Troebs}, Michael and {Vallisneri}, Michele and {Vecchio}, Alberto and {Vetrugno}, Daniele and {Vitale}, Stefano and {Volonteri}, Marta and {Wanner}, Gudrun and {Ward}, Harry and {Wass}, Peter and {Weber}, William and {Ziemer}, John and {Zweifel}, Peter},
        title = "{Laser Interferometer Space Antenna}",
      journal = {arXiv e-prints},
         year = 2017,
        month = feb,
          eid = {arXiv:1702.00786},
        pages = {arXiv:1702.00786},
          doi = {10.48550/arXiv.1702.00786},
archivePrefix = {arXiv},
       eprint = {1702.00786},
 primaryClass = {astro-ph.IM},
       adsurl = {https://ui.adsabs.harvard.edu/abs/2017arXiv170200786A}
}

@BOOK{2008gady.book.....B,
       author = {{Binney}, James and {Tremaine}, Scott},
        title = "{Galactic Dynamics: Second Edition}",
         year = 2008,
       adsurl = {https://ui.adsabs.harvard.edu/abs/2008gady.book.....B}
}

@ARTICLE{2008MNRAS.387.1416C,
       author = {{Chen}, Xuefei and {Han}, Zhanwen},
        title = "{Mass transfer from a giant star to a main-sequence companion and its contribution to long-orbital-period blue stragglers}",
      journal = {\mnras},
         year = 2008,
        month = jul,
       volume = {387},
       number = {4},
        pages = {1416-1430},
          doi = {10.1111/j.1365-2966.2008.13334.x},
archivePrefix = {arXiv},
       eprint = {0804.2294},
 primaryClass = {astro-ph},
       adsurl = {https://ui.adsabs.harvard.edu/abs/2008MNRAS.387.1416C}
}

@ARTICLE{2014A&A...563A..83C,
       author = {{Claeys}, J.~S.~W. and {Pols}, O.~R. and {Izzard}, R.~G. and {Vink}, J. and {Verbunt}, F.~W.~M.},
        title = "{Theoretical uncertainties of the Type Ia supernova rate}",
      journal = {\aap},
         year = 2014,
        month = mar,
       volume = {563},
          eid = {A83},
        pages = {A83},
          doi = {10.1051/0004-6361/201322714},
archivePrefix = {arXiv},
       eprint = {1401.2895},
 primaryClass = {astro-ph.SR},
       adsurl = {https://ui.adsabs.harvard.edu/abs/2014A&A...563A..83C}
}

@ARTICLE{2007PhRvD..75d3008C,
       author = {{Crowder}, Jeff and {Cornish}, Neil J.},
        title = "{Solution to the galactic foreground problem for LISA}",
      journal = {\prd},
         year = 2007,
        month = feb,
       volume = {75},
       number = {4},
          eid = {043008},
        pages = {043008},
          doi = {10.1103/PhysRevD.75.043008},
archivePrefix = {arXiv},
       eprint = {astro-ph/0611546},
 primaryClass = {astro-ph},
       adsurl = {https://ui.adsabs.harvard.edu/abs/2007PhRvD..75d3008C}
}

@ARTICLE{1990ApJ...358..189D,
       author = {{de Kool}, M.},
        title = "{Common Envelope Evolution and Double Cores of Planetary Nebulae}",
      journal = {\apj},
         year = 1990,
        month = jul,
       volume = {358},
        pages = {189},
          doi = {10.1086/168974},
       adsurl = {https://ui.adsabs.harvard.edu/abs/1990ApJ...358..189D}
}

@ARTICLE{2000A&A...360.1043D,
       author = {{Dewi}, J.~D.~M. and {Tauris}, T.~M.},
        title = "{On the energy equation and efficiency parameter of the common envelope evolution}",
      journal = {\aap},
         year = 2000,
        month = aug,
       volume = {360},
        pages = {1043-1051},
          doi = {10.48550/arXiv.astro-ph/0007034},
archivePrefix = {arXiv},
       eprint = {astro-ph/0007034},
 primaryClass = {astro-ph},
       adsurl = {https://ui.adsabs.harvard.edu/abs/2000A&A...360.1043D}
}

@ARTICLE{1983ApJ...268..368E,
       author = {{Eggleton}, P.~P.},
        title = "{Aproximations to the radii of Roche lobes.}",
      journal = {\apj},
         year = 1983,
        month = may,
       volume = {268},
        pages = {368-369},
          doi = {10.1086/160960},
       adsurl = {https://ui.adsabs.harvard.edu/abs/1983ApJ...268..368E}
}

@ARTICLE{2007ApJ...669.1298F,
       author = {{Fabrycky}, Daniel and {Tremaine}, Scott},
        title = "{Shrinking Binary and Planetary Orbits by Kozai Cycles with Tidal Friction}",
      journal = {\apj},
         year = 2007,
        month = nov,
       volume = {669},
       number = {2},
        pages = {1298-1315},
          doi = {10.1086/521702},
archivePrefix = {arXiv},
       eprint = {0705.4285},
 primaryClass = {astro-ph},
       adsurl = {https://ui.adsabs.harvard.edu/abs/2007ApJ...669.1298F}
}

@ARTICLE{2023A&A...669A..82L,
       author = {{Li}, Zhenwei and {Chen}, Xuefei and {Ge}, Hongwei and {Chen}, Hai-Liang and {Han}, Zhanwen},
        title = "{Influence of a mass transfer stability criterion on double white dwarf populations}",
      journal = {\aap},
         year = 2023,
        month = jan,
       volume = {669},
          eid = {A82},
        pages = {A82},
          doi = {10.1051/0004-6361/202243893},
archivePrefix = {arXiv},
       eprint = {2211.01861},
 primaryClass = {astro-ph.SR},
       adsurl = {https://ui.adsabs.harvard.edu/abs/2023A&A...669A..82L}
}

@ARTICLE{2020A&A...640A..16T,
       author = {{Toonen}, S. and {Portegies Zwart}, S. and {Hamers}, A.~S. and {Bandopadhyay}, D.},
        title = "{The evolution of stellar triples. The most common evolutionary pathways}",
      journal = {\aap},
         year = 2020,
        month = aug,
       volume = {640},
          eid = {A16},
        pages = {A16},
          doi = {10.1051/0004-6361/201936835},
archivePrefix = {arXiv},
       eprint = {2004.07848},
 primaryClass = {astro-ph.SR},
       adsurl = {https://ui.adsabs.harvard.edu/abs/2020A&A...640A..16T}
}

@ARTICLE{2010ApJ...717..724G,
       author = {{Ge}, Hongwei and {Hjellming}, Michael S. and {Webbink}, Ronald F. and {Chen}, Xuefei and {Han}, Zhanwen},
        title = "{Adiabatic Mass Loss in Binary Stars. I. Computational Method}",
      journal = {\apj},
         year = 2010,
        month = jul,
       volume = {717},
       number = {2},
        pages = {724-738},
          doi = {10.1088/0004-637X/717/2/724},
archivePrefix = {arXiv},
       eprint = {1005.3099},
 primaryClass = {astro-ph.SR},
       adsurl = {https://ui.adsabs.harvard.edu/abs/2010ApJ...717..724G}
}

@ARTICLE{2015ApJ...812...40G,
       author = {{Ge}, Hongwei and {Webbink}, Ronald F. and {Chen}, Xuefei and {Han}, Zhanwen},
        title = "{Adiabatic Mass Loss in Binary Stars. II. From Zero-age Main Sequence to the Base of the Giant Branch}",
      journal = {\apj},
         year = 2015,
        month = oct,
       volume = {812},
       number = {1},
          eid = {40},
        pages = {40},
          doi = {10.1088/0004-637X/812/1/40},
archivePrefix = {arXiv},
       eprint = {1507.04843},
 primaryClass = {astro-ph.SR},
       adsurl = {https://ui.adsabs.harvard.edu/abs/2015ApJ...812...40G}
}

@ARTICLE{2020ApJS..249....9G,
       author = {{Ge}, Hongwei and {Webbink}, Ronald F. and {Han}, Zhanwen},
        title = "{The Thermal Equilibrium Mass-loss Model and Its Applications in Binary Evolution}",
      journal = {\apjs},
         year = 2020,
        month = jul,
       volume = {249},
       number = {1},
          eid = {9},
        pages = {9},
          doi = {10.3847/1538-4365/ab98f6},
archivePrefix = {arXiv},
       eprint = {2006.00774},
 primaryClass = {astro-ph.SR},
       adsurl = {https://ui.adsabs.harvard.edu/abs/2020ApJS..249....9G}
}

@ARTICLE{2016MNRAS.459.2827H,
       author = {{Hamers}, Adrian S. and {Portegies Zwart}, Simon F.},
        title = "{Secular dynamics of hierarchical multiple systems composed of nested binaries, with an arbitrary number of bodies and arbitrary hierarchical structure. First applications to multiplanet and multistar systems}",
      journal = {\mnras},
         year = 2016,
        month = jul,
       volume = {459},
       number = {3},
        pages = {2827-2874},
          doi = {10.1093/mnras/stw784},
archivePrefix = {arXiv},
       eprint = {1511.00944},
 primaryClass = {astro-ph.SR},
       adsurl = {https://ui.adsabs.harvard.edu/abs/2016MNRAS.459.2827H}
}

@ARTICLE{2018MNRAS.476.4139H,
       author = {{Hamers}, Adrian S.},
        title = "{Secular dynamics of hierarchical multiple systems composed of nested binaries, with an arbitrary number of bodies and arbitrary hierarchical structure - II. External perturbations: flybys and supernovae}",
      journal = {\mnras},
         year = 2018,
        month = may,
       volume = {476},
       number = {3},
        pages = {4139-4161},
          doi = {10.1093/mnras/sty428},
archivePrefix = {arXiv},
       eprint = {1802.05716},
 primaryClass = {astro-ph.SR},
       adsurl = {https://ui.adsabs.harvard.edu/abs/2018MNRAS.476.4139H}
}

@ARTICLE{2020MNRAS.494.5492H,
       author = {{Hamers}, Adrian S.},
        title = "{Secular dynamics of hierarchical multiple systems composed of nested binaries, with an arbitrary number of bodies and arbitrary hierarchical structure - III. Suborbital effects: hybrid integration techniques and orbit-averaging corrections}",
      journal = {\mnras},
         year = 2020,
        month = jun,
       volume = {494},
       number = {4},
        pages = {5492-5506},
          doi = {10.1093/mnras/staa1084},
archivePrefix = {arXiv},
       eprint = {2004.08327},
 primaryClass = {astro-ph.EP},
       adsurl = {https://ui.adsabs.harvard.edu/abs/2020MNRAS.494.5492H}
}

@ARTICLE{2021MNRAS.502.4479H,
       author = {{Hamers}, Adrian S. and {Rantala}, Antti and {Neunteufel}, Patrick and {Preece}, Holly and {Vynatheya}, Pavan},
        title = "{Multiple Stellar Evolution: a population synthesis algorithm to model the stellar, binary, and dynamical evolution of multiple-star systems}",
      journal = {\mnras},
         year = 2021,
        month = apr,
       volume = {502},
       number = {3},
        pages = {4479-4512},
          doi = {10.1093/mnras/stab287},
archivePrefix = {arXiv},
       eprint = {2011.04513},
 primaryClass = {astro-ph.SR},
       adsurl = {https://ui.adsabs.harvard.edu/abs/2021MNRAS.502.4479H}
}

@ARTICLE{1998MNRAS.296.1019H,
       author = {{Han}, Zhanwen},
        title = "{The formation of double degenerates and related objects}",
      journal = {\mnras},
         year = 1998,
        month = jun,
       volume = {296},
       number = {4},
        pages = {1019-1040},
          doi = {10.1046/j.1365-8711.1998.01475.x},
       adsurl = {https://ui.adsabs.harvard.edu/abs/1998MNRAS.296.1019H}
}

@ARTICLE{1987ApJ...318..794H,
       author = {{Hjellming}, Michael S. and {Webbink}, Ronald F.},
        title = "{Thresholds for Rapid Mass Transfer in Binary System. I. Polytropic Models}",
      journal = {\apj},
         year = 1987,
        month = jul,
       volume = {318},
        pages = {794},
          doi = {10.1086/165412},
       adsurl = {https://ui.adsabs.harvard.edu/abs/1987ApJ...318..794H}
}

@ARTICLE{2000MNRAS.315..543H,
       author = {{Hurley}, Jarrod R. and {Pols}, Onno R. and {Tout}, Christopher A.},
        title = "{Comprehensive analytic formulae for stellar evolution as a function of mass and metallicity}",
      journal = {\mnras},
         year = 2000,
        month = jul,
       volume = {315},
       number = {3},
        pages = {543-569},
          doi = {10.1046/j.1365-8711.2000.03426.x},
archivePrefix = {arXiv},
       eprint = {astro-ph/0001295},
 primaryClass = {astro-ph},
       adsurl = {https://ui.adsabs.harvard.edu/abs/2000MNRAS.315..543H}
}

@ARTICLE{2002MNRAS.329..897H,
       author = {{Hurley}, Jarrod R. and {Tout}, Christopher A. and {Pols}, Onno R.},
        title = "{Evolution of binary stars and the effect of tides on binary populations}",
      journal = {\mnras},
         year = 2002,
        month = feb,
       volume = {329},
       number = {4},
        pages = {897-928},
          doi = {10.1046/j.1365-8711.2002.05038.x},
archivePrefix = {arXiv},
       eprint = {astro-ph/0201220},
 primaryClass = {astro-ph},
       adsurl = {https://ui.adsabs.harvard.edu/abs/2002MNRAS.329..897H}
}

@ARTICLE{1981A&A....99..126H,
       author = {{Hut}, P.},
        title = "{Tidal evolution in close binary systems.}",
      journal = {\aap},
         year = 1981,
        month = jun,
       volume = {99},
        pages = {126-140},
       adsurl = {https://ui.adsabs.harvard.edu/abs/1981A&A....99..126H}
}

@ARTICLE{2017MNRAS.470.1894K,
       author = {{Korol}, Valeriya and {Rossi}, Elena M. and {Groot}, Paul J. and {Nelemans}, Gijs and {Toonen}, Silvia and {Brown}, Anthony G.~A.},
        title = "{Prospects for detection of detached double white dwarf binaries with Gaia, LSST and LISA}",
      journal = {\mnras},
         year = 2017,
        month = sep,
       volume = {470},
       number = {2},
        pages = {1894-1910},
          doi = {10.1093/mnras/stx1285},
archivePrefix = {arXiv},
       eprint = {1703.02555},
 primaryClass = {astro-ph.HE},
       adsurl = {https://ui.adsabs.harvard.edu/abs/2017MNRAS.470.1894K}
}

@ARTICLE{1962AJ.....67..591K,
       author = {{Kozai}, Yoshihide},
        title = "{Secular perturbations of asteroids with high inclination and eccentricity}",
      journal = {\aj},
         year = 1962,
        month = nov,
       volume = {67},
        pages = {591-598},
          doi = {10.1086/108790},
       adsurl = {https://ui.adsabs.harvard.edu/abs/1962AJ.....67..591K}
}

@ARTICLE{2001MNRAS.322..231K,
       author = {{Kroupa}, Pavel},
        title = "{On the variation of the initial mass function}",
      journal = {\mnras},
         year = 2001,
        month = apr,
       volume = {322},
       number = {2},
        pages = {231-246},
          doi = {10.1046/j.1365-8711.2001.04022.x},
archivePrefix = {arXiv},
       eprint = {astro-ph/0009005},
 primaryClass = {astro-ph},
       adsurl = {https://ui.adsabs.harvard.edu/abs/2001MNRAS.322..231K}
}

@ARTICLE{2019MNRAS.490.5888L,
       author = {{Lamberts}, Astrid and {Blunt}, Sarah and {Littenberg}, Tyson B. and {Garrison-Kimmel}, Shea and {Kupfer}, Thomas and {Sanderson}, Robyn E.},
        title = "{Predicting the LISA white dwarf binary population in the Milky Way with cosmological simulations}",
      journal = {\mnras},
         year = 2019,
        month = dec,
       volume = {490},
       number = {4},
        pages = {5888-5903},
          doi = {10.1093/mnras/stz2834},
archivePrefix = {arXiv},
       eprint = {1907.00014},
 primaryClass = {astro-ph.HE},
       adsurl = {https://ui.adsabs.harvard.edu/abs/2019MNRAS.490.5888L}
}

@ARTICLE{2026A&A...706A.105L,
       author = {{Li}, Zhuowen and {Lu}, Xizhen and {L{\"u}}, Guoliang and {Zhu}, Chunhua and {Liu}, Helei and {Lei}, Li and {Guo}, Sufen and {He}, Xiaolong and {Beissen}, Nurzada},
        title = "{Formation of the dormant black holes with luminous companions from binary or triple systems}",
      journal = {\aap},
         year = 2026,
        month = feb,
       volume = {706},
          eid = {A105},
        pages = {A105},
          doi = {10.1051/0004-6361/202557437},
archivePrefix = {arXiv},
       eprint = {2512.04774},
 primaryClass = {astro-ph.SR},
       adsurl = {https://ui.adsabs.harvard.edu/abs/2026A&A...706A.105L}
}

@ARTICLE{2025PhRvD.112j3005L,
       author = {{Li}, Lei and {L{\"u}}, Guoliang and {Zhu}, Chunhua and {Guo}, Sufen and {Ge}, Hongwei and {Gu}, Weimin and {Li}, Zhuowen and {He}, Xiaolong},
        title = "{Explanation of the mass distribution of binary black hole mergers}",
      journal = {\prd},
         year = 2025,
        month = nov,
       volume = {112},
       number = {10},
          eid = {103005},
        pages = {103005},
          doi = {10.1103/drq9-dpy4},
archivePrefix = {arXiv},
       eprint = {2510.08231},
 primaryClass = {astro-ph.HE},
       adsurl = {https://ui.adsabs.harvard.edu/abs/2025PhRvD.112j3005L}
}

@ARTICLE{2020ApJ...890...69L,
       author = {{L{\"u}}, Guoliang and {Zhu}, Chunhua and {Wang}, Zhaojun and {Liu}, Helei and {Li}, Lin and {Xie}, Dian and {Liu}, Jinzhong},
        title = "{Possible Formation Scenarios of ZTF J153932.16+502738.8{\textemdash}A Gravitational Source Close to the Peak of LISA's Sensitivity}",
      journal = {\apj},
         year = 2020,
        month = feb,
       volume = {890},
       number = {1},
          eid = {69},
        pages = {69},
          doi = {10.3847/1538-4357/ab6bcc},
archivePrefix = {arXiv},
       eprint = {2001.03114},
 primaryClass = {astro-ph.SR},
       adsurl = {https://ui.adsabs.harvard.edu/abs/2020ApJ...890...69L}
}

@ARTICLE{1964PhRv..136.1224P,
       author = {{Peters}, P.~C.},
        title = "{Gravitational Radiation and the Motion of Two Point Masses}",
      journal = {Physical Review},
         year = 1964,
        month = nov,
       volume = {136},
       number = {4B},
        pages = {1224-1232},
          doi = {10.1103/PhysRev.136.B1224},
       adsurl = {https://ui.adsabs.harvard.edu/abs/1964PhRv..136.1224P}
}

@ARTICLE{2025A&A...700A.147W,
       author = {{Wang}, Zhijun and {L{\"u}}, Guoliang and {Zhu}, Chunhua and {Guo}, Sufen and {Liu}, Helei and {Lu}, Xizhen},
        title = "{The mixing of internal gravity waves and lithium production in intermediate-mass asymptotic giant branch stars}",
      journal = {\aap},
         year = 2025,
        month = aug,
       volume = {700},
          eid = {A147},
        pages = {A147},
          doi = {10.1051/0004-6361/202453098},
archivePrefix = {arXiv},
       eprint = {2506.20360},
 primaryClass = {astro-ph.SR},
       adsurl = {https://ui.adsabs.harvard.edu/abs/2025A&A...700A.147W}
}

@ARTICLE{1962P&SS....9..719L,
       author = {{Lidov}, M.~L.},
        title = "{The evolution of orbits of artificial satellites of planets under the action of gravitational perturbations of external bodies}",
      journal = {Planetary and Space Science},
         year = 1962,
        month = oct,
       volume = {9},
       number = {10},
        pages = {719-759},
          doi = {10.1016/0032-0633(62)90129-0},
       adsurl = {https://ui.adsabs.harvard.edu/abs/1962P&SS....9..719L}
}

@ARTICLE{1988ApJ...329..764L,
       author = {{Livio}, Mario and {Soker}, Noam},
        title = "{The Common Envelope Phase in the Evolution of Binary Stars}",
      journal = {\apj},
         year = 1988,
        month = jun,
       volume = {329},
        pages = {764},
          doi = {10.1086/166419},
       adsurl = {https://ui.adsabs.harvard.edu/abs/1988ApJ...329..764L}
}

@ARTICLE{2018MNRAS.476.2584M,
       author = {{Maoz}, Dan and {Hallakoun}, Na'ama and {Badenes}, Carles},
        title = "{The separation distribution and merger rate of double white dwarfs: improved constraints}",
      journal = {\mnras},
         year = 2018,
        month = may,
       volume = {476},
       number = {2},
        pages = {2584-2590},
          doi = {10.1093/mnras/sty339},
archivePrefix = {arXiv},
       eprint = {1801.04275},
 primaryClass = {astro-ph.SR},
       adsurl = {https://ui.adsabs.harvard.edu/abs/2018MNRAS.476.2584M}
}

@ARTICLE{2006A&A...450..681T,
       author = {{Tokovinin}, A. and {Thomas}, S. and {Sterzik}, M. and {Udry}, S.},
        title = "{Tertiary companions to close spectroscopic binaries}",
      journal = {\aap},
         year = 2006,
        month = apr,
       volume = {450},
       number = {2},
        pages = {681-693},
          doi = {10.1051/0004-6361:20054427},
archivePrefix = {arXiv},
       eprint = {astro-ph/0601518},
 primaryClass = {astro-ph},
       adsurl = {https://ui.adsabs.harvard.edu/abs/2006A&A...450..681T}
}

@ARTICLE{2008MNRAS.389..869E,
       author = {{Eggleton}, P.~P. and {Tokovinin}, A.~A.},
        title = "{A catalogue of multiplicity among bright stellar systems}",
      journal = {\mnras},
         year = 2008,
        month = sep,
       volume = {389},
       number = {2},
        pages = {869-879},
          doi = {10.1111/j.1365-2966.2008.13596.x},
archivePrefix = {arXiv},
       eprint = {0806.2878},
 primaryClass = {astro-ph},
       adsurl = {https://ui.adsabs.harvard.edu/abs/2008MNRAS.389..869E}
}

@ARTICLE{2010ApJS..190....1R,
       author = {{Raghavan}, Deepak and {McAlister}, Harold A. and {Henry}, Todd J. and {Latham}, David W. and {Marcy}, Geoffrey W. and {Mason}, Brian D. and {Gies}, Douglas R. and {White}, Russel J. and {ten Brummelaar}, Theo A.},
        title = "{A Survey of Stellar Families: Multiplicity of Solar-type Stars}",
      journal = {\apjs},
         year = 2010,
        month = sep,
       volume = {190},
       number = {1},
        pages = {1-42},
          doi = {10.1088/0067-0049/190/1/1},
archivePrefix = {arXiv},
       eprint = {1007.0414},
 primaryClass = {astro-ph.SR},
       adsurl = {https://ui.adsabs.harvard.edu/abs/2010ApJS..190....1R}
}

@ARTICLE{2026MNRAS.547ag192D,
       author = {{Di Stefano}, Rosanne and {Khwaja}, Amaan and {Kobayashi}, Chiaki},
        title = "{SCATTER common envelope formalism for triples}",
      journal = {\mnras},
         year = 2026,
        month = apr,
       volume = {547},
       number = {2},
          eid = {stag192},
        pages = {stag192},
          doi = {10.1093/mnras/stag192},
archivePrefix = {arXiv},
       eprint = {2511.04857},
 primaryClass = {astro-ph.SR},
       adsurl = {https://ui.adsabs.harvard.edu/abs/2026MNRAS.547ag192D}
}

@ARTICLE{2017ApJS..230...15M,
       author = {{Moe}, Maxwell and {Di Stefano}, Rosanne},
        title = "{Mind Your Ps and Qs: The Interrelation between Period (P) and Mass-ratio (Q) Distributions of Binary Stars}",
      journal = {\apjs},
         year = 2017,
        month = jun,
       volume = {230},
       number = {2},
          eid = {15},
        pages = {15},
          doi = {10.3847/1538-4365/aa6fb6},
archivePrefix = {arXiv},
       eprint = {1606.05347},
 primaryClass = {astro-ph.SR},
       adsurl = {https://ui.adsabs.harvard.edu/abs/2017ApJS..230...15M}
}

@ARTICLE{2001A&A...365..491N,
       author = {{Nelemans}, G. and {Yungelson}, L.~R. and {Portegies Zwart}, S.~F. and {Verbunt}, F.},
        title = "{Population synthesis for double white dwarfs . I. Close detached systems}",
      journal = {\aap},
         year = 2001,
        month = jan,
       volume = {365},
        pages = {491-507},
          doi = {10.1051/0004-6361:20000147},
archivePrefix = {arXiv},
       eprint = {astro-ph/0010457},
 primaryClass = {astro-ph},
       adsurl = {https://ui.adsabs.harvard.edu/abs/2001A&A...365..491N}
}

@ARTICLE{2004MNRAS.349..181N,
       author = {{Nelemans}, G. and {Yungelson}, L.~R. and {Portegies Zwart}, S.~F.},
        title = "{Short-period AM CVn systems as optical, X-ray and gravitational-wave sources}",
      journal = {\mnras},
         year = 2004,
        month = mar,
       volume = {349},
       number = {1},
        pages = {181-192},
          doi = {10.1111/j.1365-2966.2004.07479.x},
archivePrefix = {arXiv},
       eprint = {astro-ph/0312193},
 primaryClass = {astro-ph},
       adsurl = {https://ui.adsabs.harvard.edu/abs/2004MNRAS.349..181N}
}

@INPROCEEDINGS{1976IAUS...73...75P,
       author = {{Paczynski}, B.},
        title = "{Common Envelope Binaries}",
    booktitle = {Structure and Evolution of Close Binary Systems},
         year = 1976,
       editor = {{Eggleton}, Peter and {Mitton}, Simon and {Whelan}, John},
       series = {IAU Symposium},
       volume = {73},
        month = jan,
        pages = {75},
       adsurl = {https://ui.adsabs.harvard.edu/abs/1976IAUS...73...75P}
}

@ARTICLE{2011AN....332..450P,
       author = {{Pavlovskii}, K. and {Ivanova}, N.},
        title = "{Adiabatic mass loss from stars with convective envelopes: the structural response method}",
      journal = {Astronomische Nachrichten},
         year = 2011,
        month = may,
       volume = {332},
        pages = {450},
          doi = {10.1002/asna.201111558},
       adsurl = {https://ui.adsabs.harvard.edu/abs/2011AN....332..450P}
}

@ARTICLE{2015MNRAS.449.4415P,
       author = {{Pavlovskii}, K. and {Ivanova}, N.},
        title = "{Mass transfer from giant donors}",
      journal = {\mnras},
         year = 2015,
        month = jun,
       volume = {449},
       number = {4},
        pages = {4415-4427},
          doi = {10.1093/mnras/stv619},
archivePrefix = {arXiv},
       eprint = {1410.5109},
 primaryClass = {astro-ph.SR},
       adsurl = {https://ui.adsabs.harvard.edu/abs/2015MNRAS.449.4415P}
}

@ARTICLE{1996A&A...309..179P,
       author = {{Portegies Zwart}, S.~F. and {Verbunt}, F.},
        title = "{Population synthesis of high-mass binaries.}",
      journal = {\aap},
         year = 1996,
        month = may,
       volume = {309},
        pages = {179-196},
       adsurl = {https://ui.adsabs.harvard.edu/abs/1996A&A...309..179P}
}

@ARTICLE{2023RAA....23b5021Z,
       author = {{Zhu}, Chun-Hua and {L{\"u}}, Guo-Liang and {Lu}, Xi-Zhen and {He}, Jie},
        title = "{Formation and Destiny of White Dwarf and Be Star Binaries}",
      journal = {Research in Astronomy and Astrophysics},
         year = 2023,
        month = feb,
       volume = {23},
       number = {2},
          eid = {025021},
        pages = {025021},
          doi = {10.1088/1674-4527/acafc7},
archivePrefix = {arXiv},
       eprint = {2304.02615},
 primaryClass = {astro-ph.SR},
       adsurl = {https://ui.adsabs.harvard.edu/abs/2023RAA....23b5021Z}
}

@ARTICLE{2023ApJ...950....9R,
       author = {{Rajamuthukumar}, Abinaya Swaruba and {Hamers}, Adrian S. and {Neunteufel}, Patrick and {Pakmor}, R{\"u}diger and {de Mink}, Selma E.},
        title = "{Triple Evolution: An Important Channel in the Formation of Type Ia Supernovae}",
      journal = {\apj},
         year = 2023,
        month = jun,
       volume = {950},
       number = {1},
          eid = {9},
        pages = {9},
          doi = {10.3847/1538-4357/acc86c},
archivePrefix = {arXiv},
       eprint = {2211.04463},
 primaryClass = {astro-ph.SR},
       adsurl = {https://ui.adsabs.harvard.edu/abs/2023ApJ...950....9R}
}

@ARTICLE{2025PhRvD.111j3004L,
       author = {{Lu}, Xizhen and {Zhu}, Chunhua and {L{\"u}}, Guoliang and {Guo}, Sufen and {Li}, Zhuowen and {Zhao}, Gang},
        title = "{Impact of neutrino magnetic moments on the evolution of the helium flash and lithium-rich red clump stars}",
      journal = {\prd},
         year = 2025,
        month = may,
       volume = {111},
       number = {10},
          eid = {103004},
        pages = {103004},
          doi = {10.1103/PhysRevD.111.103004},
archivePrefix = {arXiv},
       eprint = {2504.05671},
 primaryClass = {astro-ph.SR},
       adsurl = {https://ui.adsabs.harvard.edu/abs/2025PhRvD.111j3004L}
}

@ARTICLE{2025PhRvD.111d3035W,
       author = {{Wang}, Hao and {Zhu}, Chunhua and {L{\"u}}, Guoliang and {Li}, Lin and {Liu}, Helei and {Guo}, Sufen and {Lu}, Xizhen},
        title = "{Neutrino luminosity and the energy spectrum of a nova outburst}",
      journal = {\prd},
         year = 2025,
        month = feb,
       volume = {111},
       number = {4},
          eid = {043035},
        pages = {043035},
          doi = {10.1103/PhysRevD.111.043035},
archivePrefix = {arXiv},
       eprint = {2501.13259},
 primaryClass = {astro-ph.SR},
       adsurl = {https://ui.adsabs.harvard.edu/abs/2025PhRvD.111d3035W}
}

@ARTICLE{2025A&A...704A.156R,
       author = {{Rajamuthukumar}, Abinaya Swaruba and {Korol}, Valeriya and {Stegmann}, Jakob and {Preece}, Holly and {Pakmor}, R{\"u}diger and {Justham}, Stephen and {Toonen}, Silvia and {de Mink}, Selma E.},
        title = "{The role of triple evolution in the formation of LISA double white dwarfs}",
      journal = {\aap},
         year = 2025,
        month = dec,
       volume = {704},
          eid = {A156},
        pages = {A156},
          doi = {10.1051/0004-6361/202554277},
archivePrefix = {arXiv},
       eprint = {2502.09607},
 primaryClass = {astro-ph.SR},
       adsurl = {https://ui.adsabs.harvard.edu/abs/2025A&A...704A.156R}
}

@ARTICLE{1991ARA&A..29..129R,
       author = {{Rana}, Narayan C.},
        title = "{Chemical evolution of the Galaxy.}",
      journal = {\araa},
         year = 1991,
        month = jan,
       volume = {29},
        pages = {129-162},
          doi = {10.1146/annurev.aa.29.090191.001021},
       adsurl = {https://ui.adsabs.harvard.edu/abs/1991ARA&A..29..129R}
}

@ARTICLE{2020MNRAS.492.4131R,
       author = {{Rantala}, Antti and {Pihajoki}, Pauli and {Mannerkoski}, Matias and {Johansson}, Peter H. and {Naab}, Thorsten},
        title = "{MSTAR - a fast parallelized algorithmically regularized integrator with minimum spanning tree coordinates}",
      journal = {\mnras},
         year = 2020,
        month = mar,
       volume = {492},
       number = {3},
        pages = {4131-4148},
          doi = {10.1093/mnras/staa084},
archivePrefix = {arXiv},
       eprint = {2001.03180},
 primaryClass = {astro-ph.IM},
       adsurl = {https://ui.adsabs.harvard.edu/abs/2020MNRAS.492.4131R}
}

@ARTICLE{2018CQGra..35j5011R,
       author = {{Robson}, Travis and {Cornish}, Neil J.},
        title = "{Detecting Hierarchical Triples with LISA}",
      journal = {Classical and Quantum Gravity},
         year = 2018,
        month = may,
       volume = {35},
       number = {10},
          eid = {105011},
        pages = {105011},
          doi = {10.1088/1361-6382/aab96c},
archivePrefix = {arXiv},
       eprint = {1802.04321},
 primaryClass = {gr-qc},
       adsurl = {https://ui.adsabs.harvard.edu/abs/2018CQGra..35j5011R}
}

@ARTICLE{2019CQGra..36j5011R,
       author = {{Robson}, Travis and {Cornish}, Neil J. and {Liu}, Chang},
        title = "{The construction and use of LISA sensitivity curves}",
      journal = {Classical and Quantum Gravity},
         year = 2019,
        month = may,
       volume = {36},
       number = {10},
          eid = {105011},
        pages = {105011},
          doi = {10.1088/1361-6382/ab1101},
archivePrefix = {arXiv},
       eprint = {1803.01944},
 primaryClass = {astro-ph.HE},
       adsurl = {https://ui.adsabs.harvard.edu/abs/2019CQGra..36j5011R}
}

@INPROCEEDINGS{2017JPhCS.840a2024C,
       author = {{Cornish}, Neil and {Robson}, Travis},
        title = "{Galactic binary science with the new LISA design}",
    booktitle = {Journal of Physics Conference Series},
         year = 2017,
        month = may,
       volume = {840},
          eid = {012024},
        pages = {012024},
          doi = {10.1088/1742-6596/840/1/012024},
       adsurl = {https://ui.adsabs.harvard.edu/abs/2017JPhCS.840a2024C}
}

@ARTICLE{2010ApJ...717.1006R,
       author = {{Ruiter}, Ashley J. and {Belczynski}, Krzysztof and {Benacquista}, Matthew and {Larson}, Shane L. and {Williams}, Gabriel},
        title = "{The LISA Gravitational Wave Foreground: A Study of Double White Dwarfs}",
      journal = {\apj},
         year = 2010,
        month = jul,
       volume = {717},
       number = {2},
        pages = {1006-1021},
          doi = {10.1088/0004-637X/717/2/1006},
archivePrefix = {arXiv},
       eprint = {0705.3272},
 primaryClass = {astro-ph},
       adsurl = {https://ui.adsabs.harvard.edu/abs/2010ApJ...717.1006R}
}

@ARTICLE{2008ApJ...677L..55S,
       author = {{Seto}, Naoki},
        title = "{Detecting Planets around Compact Binaries with Gravitational Wave Detectors in Space}",
      journal = {\apjl},
         year = 2008,
        month = apr,
       volume = {677},
       number = {1},
        pages = {L55},
          doi = {10.1086/587785},
archivePrefix = {arXiv},
       eprint = {0802.3411},
 primaryClass = {astro-ph},
       adsurl = {https://ui.adsabs.harvard.edu/abs/2008ApJ...677L..55S}
}

@ARTICLE{2021ApJ...920...81S,
       author = {{Shao}, Yong and {Li}, Xiang-Dong},
        title = "{Population Synthesis of Black Hole Binaries with Compact Star Companions}",
      journal = {\apj},
         year = 2021,
        month = oct,
       volume = {920},
       number = {2},
          eid = {81},
        pages = {81},
          doi = {10.3847/1538-4357/ac173e},
archivePrefix = {arXiv},
       eprint = {2107.03565},
 primaryClass = {astro-ph.HE},
       adsurl = {https://ui.adsabs.harvard.edu/abs/2021ApJ...920...81S}
}

@ARTICLE{2025ApJ...978...47S,
       author = {{Shariat}, Cheyanne and {Naoz}, Smadar and {El-Badry}, Kareem and {Rodriguez}, Antonio C. and {Hansen}, Bradley M.~S. and {Angelo}, Isabel and {Stephan}, Alexander P.},
        title = "{Once a Triple, Not Always a Triple: The Evolution of Hierarchical Triples That Yield Merged Inner Binaries}",
      journal = {\apj},
         year = 2025,
        month = jan,
       volume = {978},
       number = {1},
          eid = {47},
        pages = {47},
          doi = {10.3847/1538-4357/ad944a},
archivePrefix = {arXiv},
       eprint = {2407.06257},
 primaryClass = {astro-ph.SR},
       adsurl = {https://ui.adsabs.harvard.edu/abs/2025ApJ...978...47S}
}

@ARTICLE{1997A&A...327..620S,
       author = {{Soberman}, G.~E. and {Phinney}, E.~S. and {van den Heuvel}, E.~P.~J.},
        title = "{Stability criteria for mass transfer in binary stellar evolution.}",
      journal = {\aap},
         year = 1997,
        month = nov,
       volume = {327},
        pages = {620-635},
          doi = {10.48550/arXiv.astro-ph/9703016},
archivePrefix = {arXiv},
       eprint = {astro-ph/9703016},
 primaryClass = {astro-ph},
       adsurl = {https://ui.adsabs.harvard.edu/abs/1997A&A...327..620S}
}

@ARTICLE{2024MNRAS.534.1707T,
       author = {{Tang}, P. and {Eldridge}, J.~J. and {Meyer}, R. and {Lamberts}, A. and {Boileau}, G. and {van Zeist}, W.~G.~J.},
        title = "{Predicting gravitational wave signals from BPASS white dwarf binary and black hole binary populations of a Milky Way-like galaxy model for LISA}",
      journal = {\mnras},
         year = 2024,
        month = nov,
       volume = {534},
       number = {3},
        pages = {1707-1728},
          doi = {10.1093/mnras/stae2154},
archivePrefix = {arXiv},
       eprint = {2405.20484},
 primaryClass = {astro-ph.GA},
       adsurl = {https://ui.adsabs.harvard.edu/abs/2024MNRAS.534.1707T}
}

@ARTICLE{2018PhRvD..98f4012R,
       author = {{Robson}, Travis and {Cornish}, Neil J. and {Tamanini}, Nicola and {Toonen}, Silvia},
        title = "{Detecting hierarchical stellar systems with LISA}",
      journal = {\prd},
         year = 2018,
        month = sep,
       volume = {98},
       number = {6},
          eid = {064012},
        pages = {064012},
          doi = {10.1103/PhysRevD.98.064012},
archivePrefix = {arXiv},
       eprint = {1806.00500},
 primaryClass = {gr-qc},
       adsurl = {https://ui.adsabs.harvard.edu/abs/2018PhRvD..98f4012R}
}

@ARTICLE{2023ApJ...945..162T,
       author = {{Thiele}, Sarah and {Breivik}, Katelyn and {Sanderson}, Robyn E. and {Luger}, Rodrigo},
        title = "{Applying the Metallicity-dependent Binary Fraction to Double White Dwarf Formation: Implications for LISA}",
      journal = {\apj},
         year = 2023,
        month = mar,
       volume = {945},
       number = {2},
          eid = {162},
        pages = {162},
          doi = {10.3847/1538-4357/aca7be},
archivePrefix = {arXiv},
       eprint = {2111.13700},
 primaryClass = {astro-ph.HE},
       adsurl = {https://ui.adsabs.harvard.edu/abs/2023ApJ...945..162T}
}

@ARTICLE{2006PhRvD..73l2001T,
       author = {{Timpano}, Seth E. and {Rubbo}, Louis J. and {Cornish}, Neil J.},
        title = "{Characterizing the galactic gravitational wave background with LISA}",
      journal = {\prd},
         year = 2006,
        month = jun,
       volume = {73},
       number = {12},
          eid = {122001},
        pages = {122001},
          doi = {10.1103/PhysRevD.73.122001},
archivePrefix = {arXiv},
       eprint = {gr-qc/0504071},
 primaryClass = {gr-qc},
       adsurl = {https://ui.adsabs.harvard.edu/abs/2006PhRvD..73l2001T}
}

@ARTICLE{2018ApJS..235....6T,
       author = {{Tokovinin}, Andrei},
        title = "{The Updated Multiple Star Catalog}",
      journal = {\apjs},
         year = 2018,
        month = mar,
       volume = {235},
       number = {1},
          eid = {6},
        pages = {6},
          doi = {10.3847/1538-4365/aaa1a5},
archivePrefix = {arXiv},
       eprint = {1712.04750},
 primaryClass = {astro-ph.SR},
       adsurl = {https://ui.adsabs.harvard.edu/abs/2018ApJS..235....6T}
}

@ARTICLE{1997MNRAS.291..732T,
       author = {{Tout}, Christopher A. and {Aarseth}, Sverre J. and {Pols}, Onno R. and {Eggleton}, Peter P.},
        title = "{Rapid binary star evolution for N-body simulations and population synthesis}",
      journal = {\mnras},
         year = 1997,
        month = nov,
       volume = {291},
       number = {4},
        pages = {732-748},
          doi = {10.1093/mnras/291.4.732},
       adsurl = {https://ui.adsabs.harvard.edu/abs/1997MNRAS.291..732T}
}

@ARTICLE{2022MNRAS.516.4146V,
       author = {{Vynatheya}, Pavan and {Hamers}, Adrian S. and {Mardling}, Rosemary A. and {Bellinger}, Earl P.},
        title = "{Algebraic and machine learning approach to hierarchical triple-star stability}",
      journal = {\mnras},
         year = 2022,
        month = nov,
       volume = {516},
       number = {3},
        pages = {4146-4155},
          doi = {10.1093/mnras/stac2540},
archivePrefix = {arXiv},
       eprint = {2207.03151},
 primaryClass = {astro-ph.SR},
       adsurl = {https://ui.adsabs.harvard.edu/abs/2022MNRAS.516.4146V}
}

@ARTICLE{1984ApJ...277..355W,
       author = {{Webbink}, R.~F.},
        title = "{Double white dwarfs as progenitors of R Coronae Borealis stars and type I supernovae.}",
      journal = {\apj},
         year = 1984,
        month = feb,
       volume = {277},
        pages = {355-360},
          doi = {10.1086/161701},
       adsurl = {https://ui.adsabs.harvard.edu/abs/1984ApJ...277..355W}
}

@ARTICLE{1977A&A....57..383Z,
       author = {{Zahn}, J.-P.},
        title = "{Tidal friction in close binary systems.}",
      journal = {\aap},
         year = 1977,
        month = may,
       volume = {57},
        pages = {383-394},
       adsurl = {https://ui.adsabs.harvard.edu/abs/1977A&A....57..383Z}
}

@ARTICLE{1989A&A...220..112Z,
       author = {{Zahn}, J.-P.},
        title = "{Tidal evolution of close binary stars. I - Revisiting the theory of the equilibrium tide}",
      journal = {\aap},
         year = 1989,
        month = aug,
       volume = {220},
       number = {1-2},
        pages = {112-116},
       adsurl = {https://ui.adsabs.harvard.edu/abs/1989A&A...220..112Z}
}

@ARTICLE{1997A&AS..123..305V,
       author = {{van den Hoek}, L.~B. and {Groenewegen}, M.~A.~T.},
        title = "{New theoretical yields of intermediate mass stars}",
      journal = {\aaps},
         year = 1997,
        month = jun,
       volume = {123},
        pages = {305-328},
          doi = {10.1051/aas:1997162},
       adsurl = {https://ui.adsabs.harvard.edu/abs/1997A&AS..123..305V}
}

@ARTICLE{1910AN....183..345V,
       author = {{von Zeipel}, H.},
        title = "{Sur l'application des s{\'e}ries de M. Lindstedt {\`a} l'{\'e}tude du mouvement des com{\`e}tes p{\'e}riodiques}",
      journal = {Astronomische Nachrichten},
         year = 1910,
        month = mar,
       volume = {183},
       number = {22},
        pages = {345},
          doi = {10.1002/asna.19091832202},
       adsurl = {https://ui.adsabs.harvard.edu/abs/1910AN....183..345V}
}

@ARTICLE{2003MNRAS.341..662C,
       author = {{Chen}, Xuefei and {Han}, Zhanwen},
        title = "{Low- and intermediate-mass close binary evolution and the initial-final mass relation - III. Conservative case with convective overshooting and non-conservative case without overshooting}",
      journal = {\mnras},
         year = 2003,
        month = may,
       volume = {341},
       number = {2},
        pages = {662-668},
          doi = {10.1046/j.1365-8711.2003.06449.x},
       adsurl = {https://ui.adsabs.harvard.edu/abs/2003MNRAS.341..662C}
}

@ARTICLE{2024arXiv240207571C,
       author = {{Colpi}, Monica and {Danzmann}, Karsten and {Hewitson}, Martin and {Holley-Bockelmann}, Kelly and {Jetzer}, Philippe and {Nelemans}, Gijs and {Petiteau}, Antoine and {Shoemaker}, David and {Sopuerta}, Carlos and {Stebbins}, Robin and {Tanvir}, Nial and {Ward}, Henry and {Weber}, William Joseph and {Thorpe}, Ira and {Daurskikh}, Anna and {Deep}, Atul and {Fern{\'a}ndez N{\'u}{\~n}ez}, Ignacio and {Garc{\'\i}a Marirrodriga}, C{\'e}sar and {Gehler}, Martin and {Halain}, Jean-Philippe and {Jennrich}, Oliver and {Lammers}, Uwe and {Larra{\~n}aga}, Jonan and {Lieser}, Maike and {L{\"u}tzgendorf}, Nora and {Martens}, Waldemar and {Mondin}, Linda and {Piris Ni{\~n}o}, Ana and {Amaro-Seoane}, Pau and {Arca Sedda}, Manuel and {Auclair}, Pierre and {Babak}, Stanislav and {Baghi}, Quentin and {Baibhav}, Vishal and {Baker}, Tessa and {Bayle}, Jean-Baptiste and {Berry}, Christopher and {Berti}, Emanuele and {Boileau}, Guillaume and {Bonetti}, Matteo and {Brito}, Richard and {Buscicchio}, Riccardo and {Calcagni}, Gianluca and {Capelo}, Pedro R. and {Caprini}, Chiara and {Caputo}, Andrea and {Castelli}, Eleonora and {Chen}, Hsin-Yu and {Chen}, Xian and {Chua}, Alvin and {Davies}, Gareth and {Derdzinski}, Andrea and {Domcke}, Valerie Fiona and {Doneva}, Daniela and {Dvorkin}, Irna and {Mar{\'\i}a Ezquiaga}, Jose and {Gair}, Jonathan and {Haiman}, Zoltan and {Harry}, Ian and {Hartwig}, Olaf and {Hees}, Aurelien and {Heffernan}, Anna and {Husa}, Sascha and {Izquierdo-Villalba}, David and {Karnesis}, Nikolaos and {Klein}, Antoine and {Korol}, Valeriya and {Korsakova}, Natalia and {Kupfer}, Thomas and {Laghi}, Danny and {Lamberts}, Astrid and {Larson}, Shane and {Le Jeune}, Maude and {Lewicki}, Marek and {Littenberg}, Tyson and {Madge}, Eric and {Mangiagli}, Alberto and {Marsat}, Sylvain and {Vilchez}, Ivan Martin and {Maselli}, Andrea and {Mathews}, Josh and {van de Meent}, Maarten and {Muratore}, Martina and {Nardini}, Germano and {Pani}, Paolo and {Peloso}, Marco and {Pieroni}, Mauro and {Pound}, Adam and {Quelquejay-Leclere}, Hippolyte and {Ricciardone}, Angelo and {Rossi}, Elena Maria and {Sartirana}, Andrea and {Savalle}, Etienne and {Sberna}, Laura and {Sesana}, Alberto and {Shoemaker}, Deirdre and {Slutsky}, Jacob and {Sotiriou}, Thomas and {Speri}, Lorenzo and {Staab}, Martin and {Steer}, Dani{\`e}le and {Tamanini}, Nicola and {Tasinato}, Gianmassimo and {Torrado}, Jesus and {Torres-Orjuela}, Alejandro and {Toubiana}, Alexandre and {Vallisneri}, Michele and {Vecchio}, Alberto and {Volonteri}, Marta and {Yagi}, Kent and {Zwick}, Lorenz},
        title = "{LISA Definition Study Report}",
      journal = {arXiv e-prints},
         year = 2024,
        month = feb,
          eid = {arXiv:2402.07571},
        pages = {arXiv:2402.07571},
          doi = {10.48550/arXiv.2402.07571},
archivePrefix = {arXiv},
       eprint = {2402.07571},
 primaryClass = {astro-ph.CO},
       adsurl = {https://ui.adsabs.harvard.edu/abs/2024arXiv240207571C}
}

@ARTICLE{2024ApJ...969...68H,
       author = {{Heintz}, Tyler M. and {Hermes}, J.~J. and {Tremblay}, P.-E. and {Ould Rouis}, Lou Baya and {Reding}, Joshua S. and {Kaiser}, B.~C. and {van Saders}, Jennifer L.},
        title = "{A Test of Spectroscopic Age Estimates of White Dwarfs Using Wide WD+WD Binaries}",
      journal = {\apj},
         year = 2024,
        month = jul,
       volume = {969},
       number = {1},
          eid = {68},
        pages = {68},
          doi = {10.3847/1538-4357/ad479b},
archivePrefix = {arXiv},
       eprint = {2405.02423},
 primaryClass = {astro-ph.SR},
       adsurl = {https://ui.adsabs.harvard.edu/abs/2024ApJ...969...68H}
}

@ARTICLE{2021PhRvD.104d3019K,
       author = {{Karnesis}, Nikolaos and {Babak}, Stanislav and {Pieroni}, Mauro and {Cornish}, Neil and {Littenberg}, Tyson},
        title = "{Characterization of the stochastic signal originating from compact binary populations as measured by LISA}",
      journal = {\prd},
         year = 2021,
        month = aug,
       volume = {104},
       number = {4},
          eid = {043019},
        pages = {043019},
          doi = {10.1103/PhysRevD.104.043019},
archivePrefix = {arXiv},
       eprint = {2103.14598},
 primaryClass = {astro-ph.IM},
       adsurl = {https://ui.adsabs.harvard.edu/abs/2021PhRvD.104d3019K}
}

@ARTICLE{2022MNRAS.511.5936K,
       author = {{Korol}, Valeriya and {Hallakoun}, Na'ama and {Toonen}, Silvia and {Karnesis}, Nikolaos},
        title = "{Observationally driven Galactic double white dwarf population for LISA}",
      journal = {\mnras},
         year = 2022,
        month = apr,
       volume = {511},
       number = {4},
        pages = {5936-5947},
          doi = {10.1093/mnras/stac415},
archivePrefix = {arXiv},
       eprint = {2109.10972},
 primaryClass = {astro-ph.HE},
       adsurl = {https://ui.adsabs.harvard.edu/abs/2022MNRAS.511.5936K}
}

@ARTICLE{2018MNRAS.480..302K,
       author = {{Kupfer}, T. and {Korol}, V. and {Shah}, S. and {Nelemans}, G. and {Marsh}, T.~R. and {Ramsay}, G. and {Groot}, P.~J. and {Steeghs}, D.~T.~H. and {Rossi}, E.~M.},
        title = "{LISA verification binaries with updated distances from Gaia Data Release 2}",
      journal = {\mnras},
         year = 2018,
        month = oct,
       volume = {480},
       number = {1},
        pages = {302-309},
          doi = {10.1093/mnras/sty1545},
archivePrefix = {arXiv},
       eprint = {1805.00482},
 primaryClass = {astro-ph.SR},
       adsurl = {https://ui.adsabs.harvard.edu/abs/2018MNRAS.480..302K}
}

@TECHREPORT{2018LISA...SRD...ESA,
       author = {{LISA Science Study Team}},
        title = "{LISA Science Requirements Document}",
  institution = {European Space Agency},
         year = 2018,
        number = {ESA-LISA-EST-MIS-RS-001},
       adsurl = {https://ui.adsabs.harvard.edu/abs/2018LISA...SRD...ESA}
}

@ARTICLE{2019MNRAS.483.5518K,
       author = {{Korol}, Valeriya and {Rossi}, Elena M. and {Barausse}, Enrico},
        title = "{A multimessenger study of the Milky Way's stellar disc and bulge with LISA, Gaia, and LSST}",
      journal = {\mnras},
         year = 2019,
        month = mar,
       volume = {483},
       number = {4},
        pages = {5518-5533},
          doi = {10.1093/mnras/sty3440},
       adsurl = {https://ui.adsabs.harvard.edu/abs/2019MNRAS.483.5518K}
}

@ARTICLE{2020ApJ...901....4B,
       author = {{Breivik}, Katelyn and {Mingarelli}, Chiara M.~F. and {Larson}, Shane L.},
        title = "{Constraining Galactic Structure with the LISA White Dwarf Foreground}",
      journal = {\apj},
         year = 2020,
        month = sep,
       volume = {901},
       number = {1},
          eid = {4},
        pages = {4},
          doi = {10.3847/1538-4357/abab99},
       adsurl = {https://ui.adsabs.harvard.edu/abs/2020ApJ...901....4B}
}

@ARTICLE{2021A&A...651A.100O,
       author = {{Olejak}, A. and {Belczynski}, K. and {Ivanova}, N.},
        title = "{Impact of common envelope development criteria on the formation of LIGO/Virgo sources}",
      journal = {\aap},
         year = 2021,
        month = jul,
       volume = {651},
          eid = {A100},
        pages = {A100},
          doi = {10.1051/0004-6361/202140520},
       adsurl = {https://ui.adsabs.harvard.edu/abs/2021A&A...651A.100O}
}

@ARTICLE{2019MNRAS.490.3740N,
       author = {{Neijssel}, Coenraad J. and {Vigna-G{\'o}mez}, Alejandro and {Stevenson}, Simon and {Barrett}, Jim W. and {Gaebel}, Sebastian M. and {Broekgaarden}, Floor S. and {de Mink}, Selma E. and {Sz{\'e}csi}, Dorottya and {Vinciguerra}, Serena and {Mandel}, Ilya},
        title = "{The effect of the metallicity-specific star formation history on double compact object mergers}",
      journal = {\mnras},
         year = 2019,
        month = dec,
       volume = {490},
       number = {3},
        pages = {3740-3759},
          doi = {10.1093/mnras/stz2840},
archivePrefix = {arXiv},
       eprint = {1906.08136},
 primaryClass = {astro-ph.SR},
       adsurl = {https://ui.adsabs.harvard.edu/abs/2019MNRAS.490.3740N}
}

@ARTICLE{2012ApJ...758..131N,
       author = {{Nissanke}, Samaya and {Vallisneri}, Michele and {Nelemans}, Gijs and {Prince}, Thomas A.},
        title = "{Gravitational-wave Emission from Compact Galactic Binaries}",
      journal = {\apj},
         year = 2012,
        month = oct,
       volume = {758},
       number = {2},
          eid = {131},
        pages = {131},
          doi = {10.1088/0004-637X/758/2/131},
archivePrefix = {arXiv},
       eprint = {1201.4613},
 primaryClass = {astro-ph.GA},
       adsurl = {https://ui.adsabs.harvard.edu/abs/2012ApJ...758..131N}
}

@ARTICLE{2013ARA&A..51..269D,
       author = {{Duch{\^e}ne}, Gaspard and {Kraus}, Adam},
        title = "{Stellar Multiplicity}",
      journal = {\araa},
         year = 2013,
        month = aug,
       volume = {51},
       number = {1},
        pages = {269-310},
          doi = {10.1146/annurev-astro-081710-102602},
       adsurl = {https://ui.adsabs.harvard.edu/abs/2013ARA%26A..51..269D}
}

@ARTICLE{2012Sci...337..444S,
       author = {{Sana}, H. and {de Mink}, S.~E. and {de Koter}, A. and {Langer}, N. and {Evans}, C.~J. and {Gieles}, M. and {Gosset}, E. and {Izzard}, R.~G. and {Le Bouquin}, J. -B. and {Schneider}, F.~R.~N.},
        title = "{Binary Interaction Dominates the Evolution of Massive Stars}",
      journal = {Science},
         year = 2012,
        month = jul,
       volume = {337},
       number = {6093},
        pages = {444},
          doi = {10.1126/science.1223344},
archivePrefix = {arXiv},
       eprint = {1207.6397},
 primaryClass = {astro-ph.SR},
       adsurl = {https://ui.adsabs.harvard.edu/abs/2012Sci...337..444S}
}

@ARTICLE{2016ARA&A..54..441N,
       author = {{Naoz}, Smadar},
        title = "{The Eccentric Kozai-Lidov Effect and Its Applications}",
      journal = {\araa},
         year = 2016,
        month = sep,
       volume = {54},
        pages = {441-489},
          doi = {10.1146/annurev-astro-081915-023315},
archivePrefix = {arXiv},
       eprint = {1601.07175},
 primaryClass = {astro-ph.EP},
       adsurl = {https://ui.adsabs.harvard.edu/abs/2016ARA%26A..54..441N}
}

@ARTICLE{2013A&ARv..21...59I,
       author = {{Ivanova}, Natalia and {Justham}, Stephen and {Chen}, Xuefei and {De Marco}, Orsola and {Fryer}, Chris L. and {Gaburov}, Evghenii and {Ge}, Hongwei and {Glebbeek}, Evert and {Han}, Zhanwen and {Li}, Xiang-Dong and {Lu}, Guoliang and {Marsh}, Thomas and {Podsiadlowski}, Philipp and {Potter}, Adrian and {Soker}, Noam and {Taam}, Ronald and {Tauris}, Thomas M. and {van den Heuvel}, Evert P.~J. and {Webbink}, Ronald F.},
        title = "{Common envelope evolution: where we stand and how we can move forward}",
      journal = {\aapr},
         year = 2013,
        month = feb,
       volume = {21},
          eid = {59},
        pages = {59},
          doi = {10.1007/s00159-013-0059-2},
archivePrefix = {arXiv},
       eprint = {1209.4302},
 primaryClass = {astro-ph.HE},
       adsurl = {https://ui.adsabs.harvard.edu/abs/2013A%26ARv..21...59I}
}

@ARTICLE{2014LRR....17....3P,
       author = {{Postnov}, Konstantin A. and {Yungelson}, Lev R.},
        title = "{The Evolution of Compact Binary Star Systems}",
      journal = {Living Reviews in Relativity},
         year = 2014,
        month = may,
       volume = {17},
       number = {1},
          eid = {3},
        pages = {3},
          doi = {10.12942/lrr-2014-3},
archivePrefix = {arXiv},
       eprint = {1403.4754},
 primaryClass = {astro-ph.HE},
       adsurl = {https://ui.adsabs.harvard.edu/abs/2014LRR....17....3P}
}

@ARTICLE{2012A&A...546A..70T,
       author = {{Toonen}, S. and {Nelemans}, G. and {Portegies Zwart}, S.},
        title = "{Supernova Type Ia progenitors from merging double white dwarfs. Using a new population synthesis model}",
      journal = {\aap},
         year = 2012,
        month = oct,
       volume = {546},
          eid = {A70},
        pages = {A70},
          doi = {10.1051/0004-6361/201218966},
archivePrefix = {arXiv},
       eprint = {1208.6446},
 primaryClass = {astro-ph.SR},
       adsurl = {https://ui.adsabs.harvard.edu/abs/2012A%26A...546A..70T}
}

@ARTICLE{2005ApJ...633L..33S,
       author = {{Seto}, Naoki},
        title = "{LISA's Angular Resolution and the Verification Binary}",
      journal = {\apjl},
         year = 2005,
        month = nov,
       volume = {633},
       number = {1},
        pages = {L33-L36},
          doi = {10.1086/497693},
archivePrefix = {arXiv},
       eprint = {astro-ph/0507336},
 primaryClass = {astro-ph},
       adsurl = {https://ui.adsabs.harvard.edu/abs/2005ApJ...633L..33S}
}

@ARTICLE{2006CQGra..23S.809S,
       author = {{Stroeer}, Alexander and {Vecchio}, Alberto},
        title = "{The LISA verification binaries}",
      journal = {Classical and Quantum Gravity},
         year = 2006,
        month = oct,
       volume = {23},
       number = {19},
        pages = {S809-S817},
          doi = {10.1088/0264-9381/23/19/S19},
archivePrefix = {arXiv},
       eprint = {astro-ph/0605227},
 primaryClass = {astro-ph},
       adsurl = {https://ui.adsabs.harvard.edu/abs/2006CQGra..23S.809S}
}

@ARTICLE{2011CQGra..28i4019M,
       author = {{Marsh}, T.~R.},
        title = "{Double white dwarfs and LISA}",
      journal = {Classical and Quantum Gravity},
         year = 2011,
        month = may,
       volume = {28},
       number = {9},
          eid = {094019},
        pages = {094019},
          doi = {10.1088/0264-9381/28/9/094019},
archivePrefix = {arXiv},
       eprint = {1101.4970},
 primaryClass = {astro-ph.SR},
       adsurl = {https://ui.adsabs.harvard.edu/abs/2011CQGra..28i4019M}
}

@ARTICLE{2023MNRAS.522.5358F,
       author = {{Finch}, Emma and {Moore}, Christopher J.},
        title = "{Identifying LISA verification binaries among Gaia sources}",
      journal = {\mnras},
         year = 2023,
        month = jul,
       volume = {522},
       number = {4},
        pages = {5358-5375},
          doi = {10.1093/mnras/stad1328},
archivePrefix = {arXiv},
       eprint = {2302.03036},
 primaryClass = {astro-ph.SR},
       adsurl = {https://ui.adsabs.harvard.edu/abs/2023MNRAS.522.5358F}
}

@ARTICLE{2020ApJ...889...49B,
       author = {{Brown}, Warren R. and {Kilic}, Mukremin and {Kosakowski}, Alekzander and {Gianninas}, A. and {Andrews}, Jeffrey J. and {Heinke}, Craig O. and {Ag{\"u}eros}, Marcel A. and {Allende Prieto}, Carlos and {Kenyon}, Scott J.},
        title = "{The ELM Survey. VIII. Ninety-eight Double White Dwarf Binaries}",
      journal = {\apj},
         year = 2020,
        month = jan,
       volume = {889},
       number = {1},
          eid = {49},
        pages = {49},
          doi = {10.3847/1538-4357/ab63cd},
archivePrefix = {arXiv},
       eprint = {1912.02474},
 primaryClass = {astro-ph.SR},
       adsurl = {https://ui.adsabs.harvard.edu/abs/2020ApJ...889...49B}
}

@ARTICLE{2019Natur.571..528B,
       author = {{Burdge}, Kevin B. and {Fuller}, Jim and {Phinney}, E.~S. and {van Roestel}, Jan and {Claret}, Antonio and {Cukanovaite}, Elena and {Bellm}, Eric C. and {Kulkarni}, Shrinivas R. and {Kupfer}, Thomas and {Prince}, Thomas A.},
        title = "{General relativistic orbital decay in a seven-minute-orbital-period eclipsing binary system}",
      journal = {\nat},
         year = 2019,
        month = jul,
       volume = {571},
       number = {7766},
        pages = {528-531},
          doi = {10.1038/s41586-019-1403-0},
archivePrefix = {arXiv},
       eprint = {1907.11291},
 primaryClass = {astro-ph.SR},
       adsurl = {https://ui.adsabs.harvard.edu/abs/2019Natur.571..528B}
}

@ARTICLE{2012ApJ...749L..11B,
       author = {{Badenes}, Carles and {Maoz}, Dan},
        title = "{The Merger Rate of Binary White Dwarfs in the Galactic Disk}",
      journal = {\apjl},
         year = 2012,
        month = apr,
       volume = {749},
       number = {1},
          eid = {L11},
        pages = {L11},
          doi = {10.1088/2041-8205/749/1/L11},
archivePrefix = {arXiv},
       eprint = {1202.5472},
 primaryClass = {astro-ph.SR},
       adsurl = {https://ui.adsabs.harvard.edu/abs/2012ApJ...749L..11B}
}

@ARTICLE{2011MNRAS.417..408R,
       author = {{Ruiter}, Ashley J. and {Belczynski}, Krzysztof and {Sim}, Stuart A. and {Hillebrandt}, Wolfgang and {Fryer}, Christopher L. and {Fink}, Markus and {Kromer}, Markus},
        title = "{Delay times and rates for Type Ia supernovae and thermonuclear explosions from double-detonation sub-Chandrasekhar mass models}",
      journal = {\mnras},
         year = 2011,
        month = oct,
       volume = {417},
       number = {1},
        pages = {408-419},
          doi = {10.1111/j.1365-2966.2011.19276.x},
archivePrefix = {arXiv},
       eprint = {1011.1407},
 primaryClass = {astro-ph.SR},
       adsurl = {https://ui.adsabs.harvard.edu/abs/2011MNRAS.417..408R}
}

@ARTICLE{2008ApJS..174..223B,
       author = {{Belczynski}, Krzysztof and {Kalogera}, Vassiliki and {Rasio}, Frederic A. and {Taam}, Ronald E. and {Zezas}, Andreas and {Bulik}, Tomasz and {Maccarone}, Thomas J. and {Ivanova}, Natalia},
        title = "{Compact Object Modeling with the StarTrack Population Synthesis Code}",
      journal = {\apjs},
         year = 2008,
        month = jan,
       volume = {174},
       number = {1},
        pages = {223-260},
          doi = {10.1086/521026},
archivePrefix = {arXiv},
       eprint = {astro-ph/0511811},
 primaryClass = {astro-ph},
       adsurl = {https://ui.adsabs.harvard.edu/abs/2008ApJS..174..223B}
}

@ARTICLE{2020ApJ...898...71B,
       author = {{Breivik}, Katelyn and {Coughlin}, Scott and {Zevin}, Michael and {Rodriguez}, Carl L. and {Kremer}, Kyle and {Ye}, Claire S. and {Andrews}, Jeff J. and {Kurkowski}, Michael and {Digman}, Matthew C. and {Larson}, Shane L. and {Rasio}, Frederic A.},
        title = "{COSMIC Variance in Binary Population Synthesis}",
      journal = {\apj},
         year = 2020,
        month = jul,
       volume = {898},
       number = {1},
          eid = {71},
        pages = {71},
          doi = {10.3847/1538-4357/ab9d85},
archivePrefix = {arXiv},
       eprint = {1911.00903},
 primaryClass = {astro-ph.SR},
       adsurl = {https://ui.adsabs.harvard.edu/abs/2020ApJ...898...71B}
}

@ARTICLE{2022ApJS..258...34R,
       author = {{Riley}, J. and {Agrawal}, P. and {Barrett}, J.~W. and {Boyett}, K.~N.~K. and {Broekgaarden}, F.~S. and {Chattopadhyay}, D. and {Gaebel}, S.~M. and {Gittins}, F. and {Hendriks}, D.~D. and {Howitt}, G. and {Justham}, S. and {Lamberts}, A. and {Mandel}, I. and {Neijssel}, C.~J. and {Riley}, T. and {Stevenson}, S. and {Vigna-G{\'o}mez}, A.},
        title = "{Rapid Stellar and Binary Population Synthesis with COMPAS}",
      journal = {\apjs},
         year = 2022,
        month = jan,
       volume = {258},
       number = {2},
          eid = {34},
        pages = {34},
          doi = {10.3847/1538-4365/ac416c},
archivePrefix = {arXiv},
       eprint = {2109.10352},
 primaryClass = {astro-ph.SR},
       adsurl = {https://ui.adsabs.harvard.edu/abs/2022ApJS..258...34R}
}

@ARTICLE{2017PASA...34...58E,
       author = {{Eldridge}, J.~J. and {Stanway}, E.~R. and {Xiao}, L. and {McClelland}, L.~A.~S. and {Taylor}, G. and {Ng}, M. and {Greis}, S.~M.~L. and {Bray}, J.~C.},
        title = "{Binary Population and Spectral Synthesis Version 2.1: Construction, Observational Verification, and New Results}",
      journal = {Publications of the Astronomical Society of Australia},
         year = 2017,
        month = nov,
       volume = {34},
          eid = {e058},
        pages = {e058},
          doi = {10.1017/pasa.2017.51},
archivePrefix = {arXiv},
       eprint = {1710.02154},
 primaryClass = {astro-ph.SR},
       adsurl = {https://ui.adsabs.harvard.edu/abs/2017PASA...34...58E}
}

@ARTICLE{2014AJ....147...86T,
       author = {{Tokovinin}, Andrei},
        title = "{From Binaries to Multiples. I. Data on F and G Dwarfs within 67 pc of the Sun}",
      journal = {\aj},
         year = 2014,
        month = apr,
       volume = {147},
       number = {4},
          eid = {86},
        pages = {86},
          doi = {10.1088/0004-6256/147/4/86},
archivePrefix = {arXiv},
       eprint = {1401.6825},
 primaryClass = {astro-ph.SR},
       adsurl = {https://ui.adsabs.harvard.edu/abs/2014AJ....147...86T}
}

@ARTICLE{2014AJ....147...87T,
       author = {{Tokovinin}, Andrei},
        title = "{From Binaries to Multiples. II. Hierarchical Multiplicity of F and G Dwarfs}",
      journal = {\aj},
         year = 2014,
        month = apr,
       volume = {147},
       number = {4},
          eid = {87},
        pages = {87},
          doi = {10.1088/0004-6256/147/4/87},
archivePrefix = {arXiv},
       eprint = {1401.6826},
 primaryClass = {astro-ph.SR},
       adsurl = {https://ui.adsabs.harvard.edu/abs/2014AJ....147...87T}
}

\end{document}